\documentclass[aps,pre,showpacs,twocolumn,superscriptaddress,amssymb]{revtex4}
\usepackage{graphicx}
\usepackage{amsmath}
\usepackage{braket}
\usepackage{amssymb}
\usepackage{makeidx}
\usepackage{makecell}
\usepackage{xcolor}
\usepackage[utf8]{inputenc}

\begin{document}

\title{Extreme matrices or how an exponential map links classical and free extreme laws}
\date{\today}
\author{Jacek Grela} 
\email{jacekgrela@gmail.com} 
\affiliation{Institute of Theoretical Physics and Mark Kac Complex Systems Research Centre, Jagiellonian University, Łojasiewicza 11, 30--348 Kraków, Poland}
\author{Maciej A. Nowak}
\email{maciej.a.nowak@uj.edu.pl} 
\affiliation{Institute of Theoretical Physics and Mark Kac Complex Systems Research Centre, Jagiellonian University, Łojasiewicza 11, 30--348 Kraków, Poland}


\begin{abstract}
Using the proposed by us \emph{thinning} approach to describe extreme matrices, we find an explicit \emph{exponentiation} formula linking classical extreme laws of Fr\'echet, Gumbel and Weibull given by Fisher-Tippet-Gnedenko classification and free extreme laws of free Fr\'echet, free Gumbel and free Weibull by Ben Arous and Voiculescu \cite{BAV2006:FREEEXTREME}. We also develop an extreme random matrix formalism, in which refined questions about extreme matrices can be answered. In particular, we demonstrate  explicit calculations for several more or less known random matrix ensembles,  providing examples of all three free extreme laws.
Finally, we present  an exact mapping, showing the  equivalence of free  extreme laws  to the Peak-Over-Threshold method in classical probability. 

\end{abstract}

\maketitle
\section{Introduction}

Extreme value theory in classical probability is the prominent application of probability calculus for several problems seeking extreme values for large number of random events. Its power comes from universality according to Fisher-Tippett-Gnedenko classification~\cite{FITIGNE} which permits only three statistical laws of extremes -- Gumbel distribution, Fr\'{e}chet distribution and Weibull distribution. Beyond applications of extreme value theory in physics in the theory of disordered systems~\cite{EXTREMEPHYS}, seminal applications include insurance, finances, hydrology, neuroscience, biology, computer science, and several others~\cite{EXTREMEOTHERS,EVT1RECORDS,EVT2RG,EVT3NN,EVT4BIO}.

Since the seminal work of \cite{WISHART}, random matrix theory became one of the most universal probabilistic tools in physics and in several multidisciplinary applications~\cite{OXFORD}. In the limit when the size of the matrix tends to infinity, random matrix theory bridges to free probability theory, which can be viewed as an operator valued (i.e. non-commutative) analogue of the classical theory of probability~\cite{DKV,SPEICHER}. Both calculi exhibit striking similarities.  Wigner's semicircle law can be viewed as an analogue of normal distribution, Mar\c cenko-Pastur spectral distribution for Wishart matrices is an analogue of Poisson distribution in classical probability calculus, and Bercovici-Pata bijection~\cite{BERCOVICIPATA} is an analogue of L\'{e}vy stable processes classification for heavy-tailed distributions. It is therefore tempting to ask the question, how far can we extend the analogies between these two formalisms?

In particular, do we have an analogue of extreme values limiting distributions for the spectra of very large random matrices, i.e. does Fisher-Tippett-Gnedenko classification exists in free probability? The positive answer to this crucial question was provided more then a decade ago by Ben Arous and Voiculescu~\cite{BAV2006:FREEEXTREME}. Using operator techniques, they have proven that free probability theory has also three limiting extreme distributions - free Gumbel, free Fr\'{e}chet and free Weibull distribution. The functional form of these limiting distributions  differs from the classical probability case, but, surprisingly, the domains of attraction are the same as in their classical counterparts.  Authors of \cite{BAV2006:FREEEXTREME} state several properties of newly found free extreme laws like their representation in terms of certain generalized Pareto distribution or relation with Balkema-de Haan-Pickands~\cite{BH1974:POTLAWS,Pic1975:POT1} classification in classical probability theory of exceedances. 

Despite various connections between classical and free calculi, an explicit link between extreme laws was lacking. In this paper we establish an \emph{exponentiation} formula between laws by comparing and contrasting existing and proposed by us approaches to free extreme laws. 


\subsection{Main results}

\begin{center}
\begin{figure*}
\includegraphics[scale=.55]{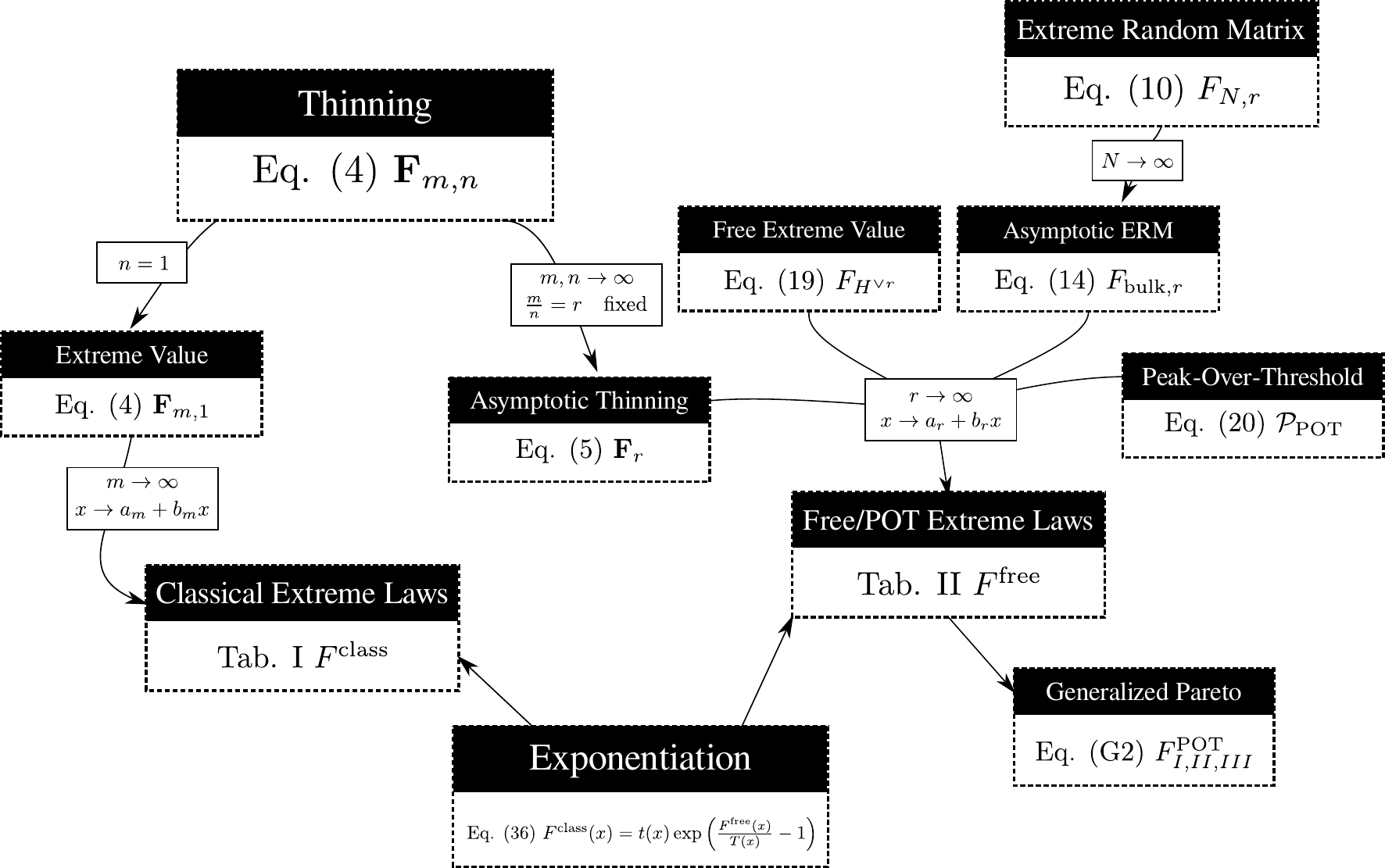}
\caption{Flow diagram of discussed frameworks in relation to families of extreme laws.}
\label{diagram}
\end{figure*}
\end{center}

Firstly, we propose two new approaches to study extremes -- \emph{thinning} method,  having root in classical extreme value theory and extreme random matrices scheme based on random matrix theory. With two previously studied frameworks due to Ben-Arous and Voiculescu (new based on free probability and old Peak-Over-Threshold statistics), we carefully establish interrelations between them summarized in Fig. \ref{diagram}. Whereas for the first three approaches equivalency is straightforward, explaining relation with Peak-Over-Threshold method is both novel and non-trivial. In the end, despite some specialization of each approach, all considered frameworks are equivalent i.e. respective cumulative distribution functions agree.

Secondly, we describe shortly the merits of the two novel approaches. \emph{Thinning} approach, in contrast to free-probabilistic method, is both intuitive and encompasses classical and free extreme events. On the other hand, extreme random matrices present a framework to which the matrix aspect of objects is accessible and presents its applicability in Fig. \ref{figfig2} when the number $r\gg N$ of extremized matrices is much larger than their sizes $N$. 

The main result \eqref{Fthin} in the thinning approach is the formula for cumulative distribution function (CDF) of a large maximum matrix obtained by extremizing $r$ large matrices:
\begin{align*}
    \mathbf{F}_{r}(x) = r (f(x) - \alpha_r) \theta(f(x) - \alpha_r),
\end{align*}
where $f$ is the CDF of a single large matrix and parameter $\alpha_r = \frac{r-1}{r}$. The result is simply a \emph{truncated} single matrix CDF $f$; see Fig. \ref{figfig1} for an example of the semicircle law.

In the extreme random matrix framework, CDF of picking the \emph{largest} out of $r$ matrices is given by Eq. \eqref{Fr} where now,  however,  the matrices are of finite size $N$:
\begin{align*}
F_{N,r}(x) = \frac{1}{N} \sum_{n=0}^{N-1} (N-n) \sum_{\substack{j_1...j_r = 0\\ j_1+...+j_r = n}}^n \prod_{l=1}^r E_N(j_l;x).
\end{align*}
The gap functions $E_N(j;x)$ are probability functions that $j$-th largest eigenvalues are greater than $x$ while $N-j$ are smaller than $x$. When $x$ is far from critical points like edges of the spectrum, two CDFs agree $\lim_{N\to \infty} F_{N,r}(x) = \mathbf{F}_{r}(x)$.

Thirdly, based on the thinning method encompassing both classical and free worlds,  we find an \emph{exponential} map \eqref{exponentiation} relating cumulative distribution functions of classical extreme laws $F^{{\rm class}}(x)$ with free extreme laws $F^{{\rm free}}(x)$:
\begin{align*}
F^{{\rm class}}(x) = t(x) \exp \left (  \frac{F^{\text{free}}(x)}{T(x)} - 1 \right ),
\end{align*}
where $t(x) = \theta(x), 1 ,1$ and $T(x) = \theta(x-1), \theta(x), \theta(x+1)$ are step functions for Fr\'{e}chet, Gumbel and Weibull classes respectively. Step function in the denominator is understood formally so as to cancel out the corresponding term in the free CDF $F^{\text{free}}$.
This proves the one-to-one correspondence between extreme laws in both probability calculi (i.e. in classical and in matricial (free) ones).

Lastly, due to operational and calculational simplicity of the thinning approach, we give several explicit examples of free extreme laws for several random matrix ensembles. As most spectral densities occurring in random matrix theory have finite supports, free Weibull category is most numerous.  Free Fr\'echet class is represented by spectral densities found through Bercovici-Pata construction being analogues of L\'{e}vy heavy-tailed distributions. Finally, 
we report in detail on one, quite exotic  example (so-called free Gaussian distribution)  found to lie within the free Gumbel class, and we comment on the link of such distribution to the  plasma physics.

\section{Approaches to extremes}
\label{frameworks}
In this section we describe status of all four approaches to extreme events and compare existing and ones proposed by us. Authors of \cite{BAV2006:FREEEXTREME} have defined free extreme values using Free Probability and also noticed  an unexpected connection to  Peak-Over-Threshold statistics. In this work we introduce two distinct but related frameworks based on random matrix theory and classical extreme value theory -- thinning method and extreme random matrices. All four approaches and their distinct properties are summarized below:
\begin{itemize}
\item Thinning method (using Statistics/Extreme Value Theory)
\begin{enumerate}
    \item works for asymptotically large matrices
    \item connects classical and free extreme laws
    \item applicable not only to eigenvalues
\end{enumerate}
\item Extreme random matrices (using Random Matrix Theory)
\begin{enumerate}
    \item works for finite matrices
    \item can address questions beyond bulk
\end{enumerate}
\item Free extreme values (using Free Probability)
\begin{enumerate}
    \item formulated in operator language
    \item not limited to matrices
\end{enumerate}
\item Peak-Over-Threshold method (using Statistics/Extreme Value Theory)
\end{itemize}

We will describe mostly first two approaches and later on compare them with the third and fourth. For ease of presentation we try to clearly delineate each approach as our goal is to highlight connections (and to large extent equivalency) between arising extreme laws. To this end, in all three descriptions we focus on a common quantity of cumulative distribution function (CDF). 

First we describe a novel approach based on Statistics and Extreme Value Theory where we focus on order statistics.  We study  the cumulative distribution function of a fraction of largest i.i.d. random variables,  which we name {\it the thinning procedure}. This approach is applicable to any random variables, not only related to random matrices. In particular, it does degenerate to the usual order statistics.

Secondly, we focus on a second new approach based on the random matrix perspective where we define en extreme cumulative distribution function $F_{N,r}$ as an average over the joint probability distribution function of the underlying ensemble of $r$ random matrices. In particular, we obtain general asymptotic results (as the matrix size $N$ goes to infinity) valid both in the bulk of the spectrum and near the (soft) spectral edge. We also investigate the limit where the number of matrices $r$ goes to infinity and a certain type of double--extreme distribution emerges.

Third point of view is based on the works \cite{BAV2006:FREEEXTREME,BAK2010:FREEEXTREME2} and has its source in Free Probability. We present a result of \cite{BG2010:MATRIXEXTREME} for the eigenvalue cumulative distribution function of maximizing $r$ Hermitian random matrices. This framework can be generalized to a general operator language. 

Lastly, we describe a Peak-Over-Threshold method which focuses on statistical study of random events only upon exceeding a certain threshold. It is somewhat unrelated to the previous ones as it is not concerned with extreme matrices per se. However, the excess distribution function $\mathcal{P}_{POT}$ is related to the CDF studied in first three approaches, and we explain this connection.

\subsection{Thinning method}
\label{thin}
Extremes of random numbers are described by \emph{order statistics}. Given a set of $m$ random variables $\{ x_1,\cdots ,x_m \} $, we rearrange them in an descending order $\{ x_{(1)},\cdots ,x_{(m)} \}$. As an example, for a set of such variables the following inequalities holds true
\begin{align*}
\begin{matrix}
x_{(1)} & \geq & x_{(2)} & \geq & \cdots & \geq & x_{(m-1)} & \geq & x_{(m)} \\
x_3 & \geq & x_{m-1} & \geq & \cdots & \geq & x_6 & \geq & x_{2}
\end{matrix} .
\end{align*}
Typically one is interested in the extreme events and studies a particular element in the ordered set $\{ x_{(1)},\cdots ,x_{(m)} \}$ -- either the largest $x_{(1)}$ or the smallest one $x_{(m)}$. One can study also the distributions of a subset of the ordered set -- the $n$ largest or smallest values. 

In all these cases, the cumulative distribution function (or CDF) for the $k$-th order statistic $x_{(k)}$ of a sample of $m$ variables is given by:
\begin{align}
\label{genp}
& \mathcal{P}^{(m)}(x_{(k)} < x) = \sum_{i=0}^{k-1} \sum_{\{ \sigma \} }\underbrace{\int_{-\infty}^x dx_{\sigma(1)} ... \int_{-\infty}^x dx_{\sigma(m-i)} }_{m-i} \nonumber \\
& \sum_{\{ \delta \} }\underbrace{\int_x^{\infty} dx_{\delta(1)} \cdots \int_x^{\infty} dx_{\delta(i)}}_{i} \mathbf{P}(x_1,...,x_{m}),
\end{align}
where $\sum_{\{ \delta \} }$ is the summation over $i$ combinations of $rN$ indices and $\sum_{\{ \sigma \} }$ is $rN-i$ combinations of the remaining $rN-i$ elements. This formula is easy to understand by positioning all particles on a line and considering only configurations where the particle with $k$-th largest position is on the left of the barrier centered at $x$ (the meaning of the condition $x_{(k)}<x$). There are $k$ possible scenarios satisfying this condition -- when the number of particles on the right side of the barrier varies from $k-1$ to $0$ which results in the summation $\sum_{i=0}^{k-1}$. Each term in the sum describes one such eventuality; the $i$ integrals $\int^{\infty}_x$ place particles on the right side of the barrier whereas the rest $rN-i$ integrals $\int_{-\infty}^x$ position the remaining ones on the left side. The only additional thing we take into account is labelling the particles which results in summation over all possible combinations $\sigma,\delta$. 

The joint probability density function (or PDF) $\mathbf{P}$ describes any set of correlated or uncorrelated random variables. In particular, for i.i.d. uncorrelated variables $\mathbf{P}(x_1,...,x_m) = \prod_{i=1}^m p(x_i)$ we find $\int_{-\infty}^y p(t) dt = f(x)$ and $\int_y^\infty p(t) dt = 1-f(x)$ which produces a well-known CDF for the $k$-th largest order statistic:
\begin{align}
\label{iid}
\mathcal{P}^{(m)}(x_{(k)} < x) = \sum_{i=0}^{k-1} \binom{m}{i} \left [1-f(x)\right ]^{i} \left [f(x) \right ]^{m-i}.
\end{align}


In the special case $k=1$, the distribution function of the largest value for $k=1$ is just $\mathcal{P}^{(m)}(x_{(1)} < x) = \left [f(x) \right ]^m$. Now we turn to describing a thinning procedure which takes a finite fraction of largest variables.

A thinning procedure applied to order statistics is to consider the following problem -- draw $m$ i.i.d. variables $\{ x_1,\cdots, x_{m} \}$ from parent PDF $p(x)$ and CDF $f(x)$, pick out the $n$ largest ones $\{ x_{(1)} \cdots x_{(n)} \}$ and look at their distribution. What will be the resulting probability density function (or PDF) and cumulative distribution function (or CDF)? We find the thinned CDF $\mathbf{F}_{m,n}$ of the $n$ largest values selected out of $m$ values as a normalized sum of first $n$ order statistics given by Eq. \eqref{iid}:
\begin{align}
\label{thinCDF}
\mathbf{F}_{m,n}(x) = \frac{1}{n} \sum_{k=1}^{n} \mathcal{P}^{(m)}(x_{(k)} < x).
\end{align}
In the i.i.d. case, we plug in Eq. \eqref{iid} into formula \eqref{thinCDF}:
\begin{align}
\label{thinCDF2}
\mathbf{F}_{m,n}(x) = \frac{1}{n} \sum_{k=0}^{n-1}(n-k) \binom{m}{k} \left [1-f(x)\right ]^{k} \left [f(x) \right ]^{m-k},
\end{align}
where we used an identity \eqref{ident0}.
Define ratio $r = \frac{m}{n}$ so we take the $m,n \to \infty$ limit such that $r$ remains fixed. An  asymptotic form of thinned CDF $\lim_{m,n\to \infty} \mathbf{F}_{m,n}(x) = \mathbf{F}_{r}(x)$ is found in App. \ref{Frhoedgeder}:
\begin{align}
\label{Fthin}
\mathbf{F}_{r}(x) = r (f(x) - \alpha_r) \theta(f(x) - \alpha_r),
\end{align}
where $\alpha_r = \frac{r-1}{r}$ and $\theta$ is a step function. It gives a CDF of a thinned population where from $m$ random elements we pick $n<m$ largest ones with ratio $r = m/n$.

Interpretation of the asymptotic thinned CDF $\mathbf{F}_r$ is clear -- picking $n$ largest values out of $m$ does not modify the shape of the parent distribution $f(x)$ but truncates it up to a point $x_*$ such that $f(x_*) = \alpha_r$. The point $x_*$ is known in statistics as the last of the $r$-quantile and gives the point where the fraction of values smaller than $x_*$ is $\alpha_r = \frac{r-1}{r}$. Importantly, since the large $n,m$ limit was taken the fraction $\alpha_r$ takes all real number between $(0,1)$. 

We stress that above discussion is purposely not restricted to matrix eigenvalues as it is applicable to general random variables. In particular, above we have addressed the simplest case of completely uncorrelated case where joint PDF factorizes which still has applications to matrices, as we will see later. Besides that, in the next section we deal with another important class of jPDFs with matrix eigenvalues as coupled/correlated random variables arising within random matrix theory.

\subsection{Extreme random matrices}
\label{extrmat}
Unlike real numbers, finding extremes in the space of matrices cannot be done easily due to lack of natural ordering. To circumvent that we instead define ordering in the space of \emph{random} matrices. This is due to an existing natural identification between the random matrix and its eigenvalues -- in almost all matrix probability distribution functions the eigenvectors completely decouple. Thus, we can disregard them completely and define extreme matrices based on eigenvalues alone. The procedure is straightforward:
\begin{enumerate}
\item Take $r$ random matrices each of size $N\times N$, collect all $N r$ eigenvalues and,
\item pick out $N$ largest ones; these form the largest/extreme random matrix representation.
\end{enumerate}
Main drawback of this definition is that the resulting extreme random matrix will contain a mixture of eigenvalues from several initial matrices.

We now introduce some useful notation and then continue describing above approach mathematically. Firstly, denote $\{ \lambda_1^{(i)} \cdots \lambda_N^{(i)} \}$ to be the set of eigenvalues of $i$-th matrix drawn from the most general joint probability density function (hereafter jPDF) for $i$-th individual matrix:
\begin{align*}
P^{(i)}_N\left (\lambda_1^{(i)},\cdots,\lambda_N^{(i)} \right ).
\end{align*}
In general, these distribution functions could differ between matrices however in what follows we consider an i.i.d. case. All $Nr$ eigenvalues are ordered in the following way:
\begin{align}
\label{trans}
\begin{matrix}
\lambda_1^{(1)} & \cdots & \lambda_N^{(1)} & \lambda_1^{(2)} & \cdots &  \lambda_1^{(r)} & \cdots & \lambda_N^{(r)} \\
\downarrow & \cdots & \downarrow & \downarrow & \cdots & \downarrow & \cdots & \downarrow \\
x_1 & \cdots & x_N &  x_{N+1} &\cdots & x_{(r-1)N+1} & \cdots & x_{rN}
\end{matrix}
\end{align}
and so the total jPDF is a product of single-matrix distributions:
\begin{align}
\label{jpdf2}
\mathbf{P} \left (x_1, \cdots , x_{rN} \right ) = \prod_{i=1}^r P^{(i)}_N \left (x_{(i-1)N + 1}, \cdots, x_{iN} \right  ).
\end{align}
Lastly we rearrange all variables $\{ x_{(1)},\cdots ,x_{(rN)} \}$ in descending order $x_{(1)}> x_{(2)} > \cdots > x_{(rN)}$ so that we pick out only the $N$ largest ones i.e. $x_{(1)}, \cdots, x_{(N)}$. 

It is important to note how such rearrangements cast our current matrix problem into a thinning framework introduced before in Sec. \ref{thin} with a special form of correlated jPDF given by Eq. \eqref{jpdf2} and substituting $n\to N, m \to rN$.

\subsubsection{Extreme cumulative distribution function $F_{N,r}$}
The CDF for the $k$-th order statistic $\mathcal{P}_r(x_{(k)} < x) = \left < \theta(x-x_{(k)}) \right >$ is given by Eq. \eqref{genp} with $m \to Nr$:
\begin{align}
\label{genp2}
& \mathcal{P}_{N,r}(x_{(k)} < x) = \sum_{i=0}^{k-1} \sum_{\{ \sigma \} }\underbrace{\int_{-\infty}^x dx_{\sigma(1)} ... \int_{-\infty}^x dx_{\sigma(rN-i)}}_{rN-i} \nonumber \\
& \sum_{\{ \delta \} }\underbrace{\int_x^{\infty} dx_{\delta(1)} ... \int_x^{\infty} dx_{\delta(i)}}_{i} \mathbf{P}(x_1,...,x_{rN}),
\end{align}
where $\sum_{\{ \delta \} }$ is the summation over $i$ combinations of $rN$ indices and $\sum_{\{ \sigma \} }$ is $rN-i$ combinations of the remaining $rN-i$ elements. Its validity was explained in previous Sec. \ref{thin}.



Since we study the density of $N$ largest eigenvalues, we define an extreme CDF as a normalized sum of terms \eqref{genp2} in analogy with Eq. \eqref{thinCDF}:
\begin{align}
\label{Frdef}
F_{N,r}(x) = \frac{1}{N} \sum_{k=1}^N \mathcal{P}_{N,r}(x_{(k)}  < x).
\end{align}
Next we make use of symmetries in the eigenvalue jPDFs and find the following form (details in App. \ref{FrApp}):
\begin{align}
\label{Fr}
F_{N,r}(x) = \frac{1}{N} \sum_{n=0}^{N-1} (N-n) \sum_{\substack{j_1...j_r = 0\\ j_1+...+j_r = n}}^n \prod_{l=1}^r E_N(j_l;x),
\end{align}
where the $k$-th gap function $E_N(k;x)$ of finding exactly $k$ (of $N$) particles in an interval $(x,+\infty)$ is defined as
\begin{align}
\label{En}
E_N(k;x) = \binom{N}{k} \int\limits_{x}^\infty d\lambda_1 ... d\lambda_k \int\limits^{x}_{-\infty} d\lambda_{k+1} ... d\lambda_N P_N,
\end{align}
with eigenvalue joint PDF $P_N(\lambda_1 \cdots \lambda_N)$. The formula \eqref{Fr} is already expressed entirely in terms of well-known objects in random matrix theory. It is exact for any value of both $N$ and $r$ however explicit forms of gap functions are not known, we consider several important cases.

For $r = 1$, Eq. \eqref{Fr} reduces to $F_{N,1}(x) = \frac{1}{N} \sum_{n=0}^{N-1} (N-n) E_N(n;x)$ and in App. \ref{F1App} we show how it is in turn given in terms of spectral density $\rho_N^{(1)}$ (or the one-point correlation function):
\begin{align}
\label{F1form}
F_{N,1}(x) = \frac{1}{N} \int_{-\infty}^x \rho_N^{(1)}(y) dy,
\end{align}
which simply means that $F_{N,1}$ is the spectral CDF. It is hardly surprising since inspecting $N$ out of $N$ largest eigenvalues should reduce exactly to quantities related with spectral density itself. 

\subsubsection{Bulk and edge limiting forms of $F_{N,r}$}
\label{FrLimits}
We turn to describe various limiting forms of extreme CDF $F_{N,r}(x)$ given by Eq. \eqref{Fr}. We address mostly cases when the argument $x$ is far from the edge of matrix spectrum (the bulk regime), when $x$ becomes close to the spectral edge (soft edge regime) and lastly comment on the double scaling limit when $r \sim N$.

\paragraph{In the bulk.}
We first evaluate $F_{N,r}$ in the bulk. To this end, in App. \ref{gapasymptotics} we calculate asymptotic form of gap function $E_{\text{bulk}} (k;x) = e^{-N(1-f(x))} \frac{[N(1-f(x))]^k}{k!},$ and plug it into Eq. \eqref{Fr} so that $F_{N,r}(x) \sim F_{\text{bulk},r}(x)$ reads:
\begin{align*}
F_{\text{bulk},r} = \frac{e^{-Nr(1-f)}}{N} \sum_{n=0}^{N-1} (N-n) \sum_{\substack{j_1...j_r = 0}}^n \frac{\left [ N(1-f) \right ]^{n}}{j_1!...j_r!} ,
\end{align*}
where for brevity we skipped the argument $x$ and the multiple sum is over indices such that $j_1+...+j_r = n$ and $f(x)$ is the asymptotic spectral CDF related to the asymptotic spectral density $\rho^{(1)}_{\text{bulk}}$. We take out the exponent and powers outside of the sums, compute the constrained multiple sum as $\sum_{j_1...j_r = 0}^n \frac{1}{j_1!} \cdots \frac{1}{j_r!} = \frac{r^n}{n!}$ and so the extreme CDF in the bulk reads:
\begin{align}
\label{Frbulk0}
F_{\text{bulk},r}(x) = e^{-N(1-f)} \frac{1}{N} \sum_{n=0}^{N-1} (N-n) \frac{1}{n!} \left [ Nr(1-f) \right ]^{n}.
\end{align}
In App. \ref{Frbulkder} we compute asymptotic form of this sum:
\begin{align}
\label{Frbulk}
F_{\text{bulk},r}(x) = r \left ( f(x) - \alpha_r \right  ) \theta \left ( f(x) - \alpha_r \right  ) ,
\end{align}
where $\alpha_r = \frac{r-1}{r}$. We stress that the formula is valid in the bulk and $f$ is the matrix CDF $f(x) = \frac{1}{N} \int_{-\infty}^x \rho_{\text{bulk}}^{(1)}(y) dy$.

We emphasize that current formula found within the matrix setup is the same as Eq. \eqref{Fthin} found in the thinning approach. 
At first glance this is a very surprising result, since in the current matrix case, we inspect CDFs of highly correlated eigenvalues while in Eq. \eqref{Fthin} we restricted the thinned approach to independent random variables! The key is in understanding the bulk region properly as an effective macroscopic picture of eigenvalues where all correlations are absorbed into the spectral density alone. 
Hence 
the macroscopic and  bulk point of view are indeed equivalent.

This relation also marks the limitations of the thinning method which holds only when the underlying eigenvalues are typical i.e. drawn from spectral PDFs. In the following we comment on a case for which this assumption does not hold.

\paragraph{Near the edge.}
Near the edge we lack an explicit formula due to strong correlations rendering the formula for gap functions intractable. Instead, we define $\lim_{N\to \infty} F_{N,r}(x_{\text{edge}} + \sigma N^{-\alpha}) = F_{\text{edge},r}(\sigma)$ and present its implicit integral representation:
\begin{align}
\label{Fredge}
F_{\text{edge},r} = \lim_{N\to \infty} \oint\limits_{\Gamma(0)} & \frac{dz}{2\pi iN z} \sum_{k=1}^{\infty} \sum_{n=0}^{k-1} \left [  \sum_{j=0}^n \frac{E_{\text{edge}}(j;\sigma) }{z^{n-j}}\right ]^r,
\end{align}
where $\Gamma(0)$ is a contour encircling $z=0$ counter-clockwise and $E_{\text{edge}}$ is given implicitly in App. \ref{gapasymptotics} by formula \eqref{Enedge}. Eq. \eqref{Fredge} is found simply from Eq. \eqref{Fr} since the contour integral is an alternative representation of the constraint $j_1+...+j_r=n$ present in the multiple sum. 

This equation is an implicit form of what we tentatively call a free Airy CDF. The name stems from the $r=1$ case where $\frac{d}{dx} F_{N,1}(x)|_{x = x_{\text{edge}} + \sigma/N^{-\alpha}}$ in general describes the spectral edge oscillations of Airy type. Hence, for general $r$ we expect a similar oscillatory pattern to emerge.

\paragraph{Large sample limit $r \to \infty$ in the bulk and the edge regimes.}
Lastly we consider the limit of large sample i.e. when $r \to \infty$ or when the number of matrices we maximize grows. In the bulk, behaviour is simple since CDF is localized inside an interval $x\in (f^{-1}(\alpha),x_{*})$ where $x_{*}$ is the rightmost edge-point in the spectrum. As we increase $r$, since $f^{-1}(\alpha) \to  f^{-1}(1) = x_{*}$, the result is a step function placed at the rightmost edge of the spectrum $x_{*}$:
\begin{align}
\label{FrbulkR}
\lim_{r\to \infty} F_{\text{bulk},r}(x) = \theta(x-x_{*}).
\end{align}
This agrees with the intuition that as we draw from an increasing pool of matrices, the result becomes a degenerate matrix with $N$ eigenvalues equal to $x_{*}$ as these are the maximal attainable eigenvalues of a matrix when studied in the bulk regime. Typically $x_*$ is also the edge-point $x_{\text{edge}}$. In the following we will consider such case.

Near the edge $x_* = x_{\text{edge}}$, we observe fluctuations of the maximal eigenvalue position and so it is no longer fixed. Moreover, we expect a natural transition point to happen around $r = N$ as then the pool of $rN \sim N^2$ eigenvalues becomes large enough so that almost surely the $N$ largest are almost all extremes from the single matrix point of view. Detailed treatment of this transition could be done through setting $r=\rho N$ in formula \eqref{Fredge}, it is intractable already for fixed and finite $r$. Instead, we offer a less rigorous but both intuitively and numerically backed approach. 

We continue with the discussion about the nature of eigenvalues as they are sifted through the maximalization procedure. For concreteness, we set $r = \rho N$ with integer $\rho\geq 1$. Then, each picked out eigenvalue has on average $\rho$ single-matrix extremes to maximize over -- we look for extremes among the (single-matrix) extremes. In other words, all non-extreme eigenvalues are almost always disregarded when looking at pool numerous enough (large $r$). This situation describes an emergent picture -- for large enough $r$, instead of picking $N$ out of $N^2\rho$ eigenvalues, we pick $N$ out of $N\rho$ \emph{extreme} eigenvalues. Crucially, it is also tractable as the joint emergent law for extreme eigenvalues $\mathbf{P}$ completely decouples and we go through all the steps in the derivation of CDF \eqref{Fr} with $F_{N,N\rho}(x_{\text{edge}} + \sigma N^{-\alpha}) \sim \bar{F}_{\text{edge},\rho}(\sigma)$:
\begin{align*}
& \bar{F}_{\text{edge},\rho}(\sigma) = \sum_{n=0}^{N-1} \frac{N-n}{N} \times \\
& \times \sum_{\substack{j_1...j_\rho = 0}}^n  \prod_{i=1}^\rho \binom{n}{j_i} \left [ 1-E_{\text{edge}}(0;\sigma) \right ]^{j_i} E_{\text{edge}}(0;\sigma)^{n-j_i}.
\end{align*}
where the summand now consists only of $ E_{\text{edge}}(0;\sigma)$ describing CDF of maximal eigenvalue Tracy--Widom formula \cite{TRACYWIDOM}, while the multiple sum is subject to a constraint $j_1+...+j_\rho = n$. We simplify above formula similarly to Eq. \eqref{Frbulk0} by using an identity $\sum_{\substack{j_1...j_\rho = 0\\ j_1+...+j_\rho = n}}^n = \binom{N\rho}{n}$ and so multiple sums reads $\bar{F}_{\text{edge},\rho}(\sigma) = \sum_{n=0}^{N-1} \frac{N-n}{N} \binom{N\rho}{n} E_{\text{edge}}(0;\sigma)^{n(\rho-1)} \left [ 1-E_{\text{edge}}(0;\sigma) \right ]^{n}$. In App. \ref{Frhoedgeder} we compute its large $N$ asymptotics:
\begin{align}
\label{Frhoedge}
\bar{F}_{\text{edge},\rho}(\sigma) = \rho \left ( E_{\text{edge}}(0;\sigma) - \alpha_\rho \right ) \theta\left ( E_{\text{edge}}(0;\sigma) - \alpha_\rho \right ),
\end{align}
where $\alpha_\rho = \frac{\rho-1}{\rho}$. The formula admits form identical to the bulk extreme CDF $F_{\text{bulk},r}$ given by Eq. \eqref{Frbulk} and thinned CDF \eqref{Fthin} upon substituting $f(x) \to E_{\text{edge}}(0,\sigma)$ and $r\to \rho$. Although it was derived for integer $\rho$, above equation can be computed for any value $\rho$. In Fig. \ref{figfig2} we present numerical experiments in the case of Gaussian Unitary Ensemble where we clearly observe regions of validity for formulas \eqref{Frbulk} and \eqref{Frhoedge}. 

It is important to note that Eq. \eqref{Frhoedge} does not give a free Tracy-Widom distribution despite consisting of a Tracy-Widom formula. A correct way of finding such law is  possible, however,  one would need to go beyond order statistics \eqref{genp2} and define global gap functions which we find as both interesting yet unsolved problem.

\subsection{Free extreme values}
Free extreme values were introduced in general operator language in Refs. \cite{BAV2006:FREEEXTREME,BAK2010:FREEEXTREME2} and the special case of extreme matrices were discussed in detail in Ref. \cite{BG2010:MATRIXEXTREME}. We review these findings in greater detail in App. \ref{appe}, while in the following we restrict ourselves to user-friendly operational definitions applicable to random matrices. 

We first define an operational definition of max operation for random Hermitian matrices $H_a \lor H_b $ -- given $2N$ eigenvalues of $H_a,H_b$, we pick out the $N$ largest eigenvalues and form the spectrum of $H_a \lor H_b$. Since a random matrix is unitarily invariant, eigenvalues alone fully specify the matrix. We state the maximal law for asymptotically large matrices given in Refs. \cite{BAV2006:FREEEXTREME,BG2010:MATRIXEXTREME}.
The main result we need from these papers is that of asymptotic eigenvalue CDF of the maximum $H_a \lor H_b$ of two random matrices $H_a,H_b$:
\begin{align}
\label{maxlaw}
F_{H_a \lor H_b}(x) = \max (0, f_{H_a}(x) + f_{H_b}(x) - 1),
\end{align}
where $f_H(x) = \int_{-\infty}^x dt \rho_{\text{bulk},H}(t)$ is the spectral CDF of the corresponding bulk PDF $\rho_{\text{bulk},H}(t) = \lim_{N\to \infty} \frac{1}{N}\left < \sum_{i=1}^N \delta(\lambda_i - t) \right >$. Single-matrix definitions of CDF and PDF are the same as to those found in previous sections \ref{thin} and \ref{extrmat} with additional subscript denoting the underlying matrix $H$. 

A special case of Eq. \eqref{maxlaw}, for a maximum of $r$ i.i.d. matrices each with eigenvalue CDF $f_H(x)$ we have:
\begin{align}
\label{maxlaw2}
F_{H^{\lor r}}(x) = \max (0, r f_H(x) - (r-1)),
\end{align}
where $H^{\lor r} = \underbrace{H \lor ... \lor H}_{r~\text{terms}}$.

\subsection{Peak-Over-Threshold method}

The method is closely related to the notion of \textit{exceedances} which arise conditioned on the event that the random variable $X$ is larger than some threshold $u$. For $ t \ge u$, the exceedance distribution function ${\rm F}_{[u]}(t)$ is then
\begin{align*}  
{\rm F}_{[u]}(t)= \mathcal{P}(X < t|X > u)& =\frac{\mathcal{P}(X < t, X>u)}{\mathcal{P}(X > u)}\\
& = \frac{f(t)-f(u)}{1-f(u)},
\end{align*}
where we used the usual definition of conditional probability $\mathcal{P}(A|B) = \mathcal{P}(A,B)/\mathcal{P}(B)$ and CDF $f(x) := \mathcal{P}(X < x)$. The Peak-Over-Threshold method (or POT) developed in Refs. \cite{Pic1975:POT1,Smi1987:POT2} in turn looks at excess distribution functions of events $X$ above some threshold $u$:
\begin{align}
\label{potdef}
\mathcal{P}_{\text{POT}}(X < u + t | X> u) = \frac{f(u+t)-f(u)}{1-f(u)}.
\end{align}
An excess of $t$ is therefore a variant of an exceedance shifted by the threshold $u$, i.e. 
${\rm F}_{[u]}(t+u)$. 

\subsection{Extreme GUE example}
\label{GUE}
In order to highlight both similarities and differences of first three frameworks we work out explicitly the case of Gaussian Unitary Ensemble (or GUE). The model is properly rescaled with jPDF given by 
\begin{align}
\label{jpdfGUE}
P(H)dH \sim \exp \left ( - \frac{N}{2} \text{Tr} H^2 \right ) dH,
\end{align}
 so that the bulk spectral density is contained within an interval $(-2,2)$ and is given by the Wigner semicircle law $\rho^{(\text{GUE})}_{\text{bulk}}(y) = \frac{1}{2\pi} \sqrt{4-y^2}$. Its CDF reads $f^{(\text{GUE})}_{\text{bulk}}(x) = \int_{-2}^x \rho^{(\text{GUE})}_{\text{bulk}}(y) dy$:
\begin{align}
\label{fGUEbulk}
f^{(\text{GUE})}_{\text{bulk}}(x) & = \frac{1}{2} + \frac{1}{4\pi} x \sqrt{4-x^2} + \frac{1}{\pi} \arcsin \frac{x}{2}.
\end{align}
\paragraph{GUE through a thinning method.}

Firstly, we look at the GUE example from the point of view of a thinning procedure. That is, we think of taking $m$ i.i.d. random variables each drawn from a CDF \eqref{fGUEbulk}. Then we look at $n$ largest variables and study the resulting quantity in the $m,n \to \infty $ limit such that $m/n=r$ remains fixed. This results in Eq. \eqref{Fthin} adapted to our example:
\begin{align}
\label{thinCDFGUE}
\mathbf{F}^{(\text{GUE})}_{r} = r \left (f^{(\text{GUE})}_{\text{bulk}}(y) - \alpha_r \right ) \theta\left (f^{(\text{GUE})}_{\text{bulk}}(y) - \alpha_r \right ),
\end{align}
where $\alpha_r = \frac{r-1}{r}$. 

\begin{center}
\begin{figure*}
\includegraphics[scale=.4]{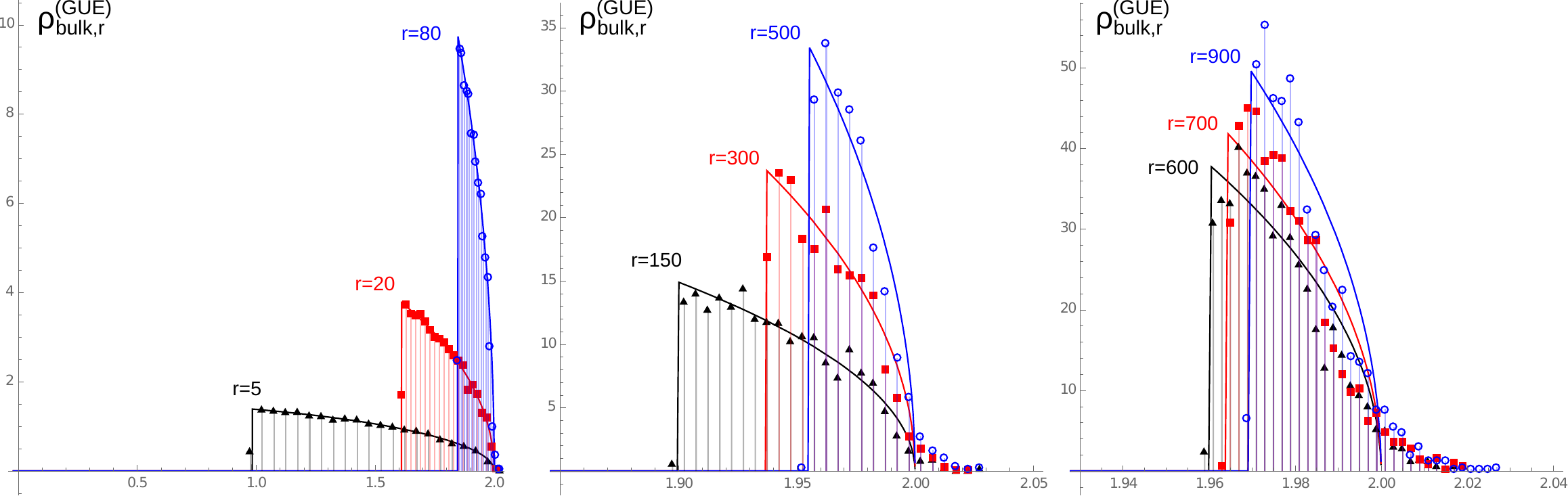}
\caption{Analytical (lines) and numerical (points) density of eigenvalues $\rho^{(\text{GUE})}_{\text{bulk},r}(x) = \frac{d}{dx} F^{(\text{GUE})}_{\text{bulk},r}(x)$ given by Eqs. \eqref{thinCDFGUE} and \eqref{FbulkGUE} where we have plotted $r$ matrices of size $500\times 500$ each drawn from an Gaussian Unitary Ensemble defined by jPDF eqref{jpdfGUE}. The change of $r$ parameter from small ($r<<N$) through medium ($r\sim N$) to large $r>N$ shows a transition where tail-like spectral features leaks out beyond the bulk boundary at $x=2$.}
\label{figfig1}
\end{figure*}
\end{center}

To check above formula we collect several numerical experiments in Fig. \ref{figfig1} (left plot) and find perfect agreement for $N=500$, $r=5,20,80$. For larger values of pool size $r$, the center and right plots of Fig. \ref{figfig1} become more concentrated around $x=2$ (notice the scales on the x-axis) i.e. in an interval $x\in \left ((f^{(\text{GUE})}_{\text{bulk}})^{-1}(\alpha_r),2 \right )$. For $\alpha_r$ sufficiently close to $1$, bulk asymptotics becomes less relevant as $(f^{(\text{GUE})}_{\text{bulk}})^{-1}(\alpha_r) \to 2$ and the thinning approach is no longer applicable. 

\paragraph{GUE as an extreme random matrix.}
We look at $r$ random matrices each drawn from jPDF \eqref{jpdfGUE} and inspect their spectra. Then we pick $N$ largest eigenvalues, form an extreme CDF given by Eq. \eqref{Frbulk}:
\begin{align}
\label{FbulkGUE}
F_{\text{bulk},r}^{(\text{GUE})} = r \left (f^{(\text{GUE})}_{\text{bulk}}(x) - \alpha \right ) \theta\left (f^{(\text{GUE})}_{\text{bulk}}(x) - \alpha \right ),
\end{align}
where $x \in (-2,2)$ and with the function $f^{(\text{GUE})}_{\text{bulk}}$ given by Eq. \eqref{fGUEbulk}. As we discussed previously, thinning method and extreme matrices in the bulk are equivalent and so are CDFs \eqref{thinCDFGUE} and \eqref{FbulkGUE}. For large pool sizes $r$ both loose its applicability and we enter the edge scaling regime.

\begin{center}
\begin{figure}[h]
\includegraphics[scale=.68]{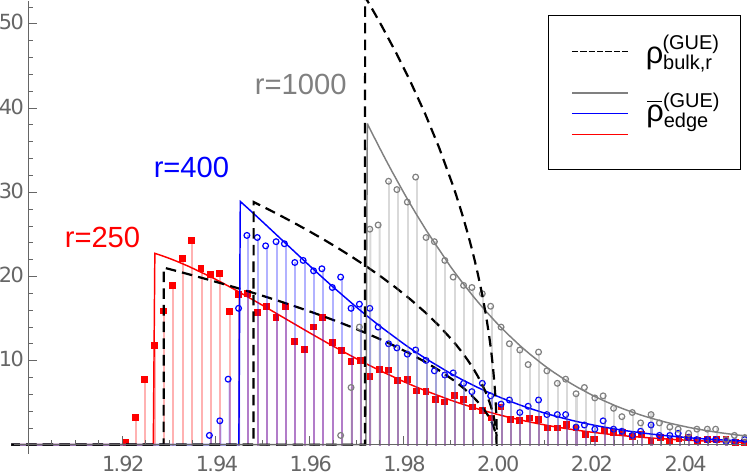}
\caption{Numerical and analytic plots of eigenvalues comprising of solid lines for the edge density and dashed black lines for bulk density and points for the histograms. Density of eigenvalues in bulk is given by $\rho^{(\text{GUE})}_{\text{bulk},r}(x) = \frac{d}{dx} F^{(\text{GUE})}_{\text{bulk},r}(x)$ and Eqs. \eqref{thinCDFGUE}, \eqref{FbulkGUE} whereas the edge formula holds with CDF given by Eq. \eqref{FrhoedgeGUE}. Histograms are plotted for matrices of size $100\times 100$ each drawn from the Gaussian Unitary Ensemble defined by jPDF \eqref{jpdfGUE}. We present only large $r$ parameters $r>N$ where the spectrum deviates considerably from the bulk description and is in turn described well by the edge formula.}
\label{figfig2}
\end{figure}
\end{center}

For larger values of $r$, the center and right plots of Fig. \ref{figfig1} become more concentrated around $x=2$, the soft-edge asymptotics become dominant when $\alpha_r$ is such that $2-(f^{(\text{GUE})}_{\text{bulk}})^{-1}(\alpha_r) \sim N^{-2/3}$ or when $ r \sim N$. In Fig. \ref{figfig1} we notice this transition through appearance of tail-like features missed altogether by the bulk formula and captured by implicit formula Eq. \eqref{Fredge}. We pass to values $r=500,600,700,900$ with $N=500$ and enter into region of increasing tails reaching beyond the bulk boundary at $x=2$ where we have at our disposal an explicit formula for the extreme CDF given by Eq. \eqref{Frhoedge}. In our example of GUE it reads:
\begin{align}
\label{FrhoedgeGUE}
\bar{F}^{(\text{GUE})}_{\text{edge},\rho}(\sigma) = \rho (E(0;\sigma) - \alpha_\rho) \theta(E(0;\sigma) - \alpha_\rho),
\end{align}
where $E(0;\sigma)$ is the CDF of $\beta=2$ Tracy-Widom law and $\alpha_\rho = \frac{\rho-1}{\rho}$. We tentatively name it a \emph{double} extreme law as it sieves out extremes among extremes. In Fig. \ref{figfig2} we show how well the large $r$ formula \eqref{FrhoedgeGUE} fits the simulations and also juxtapose it with the bulk formula \eqref{FbulkGUE}. 
%
%
%

\paragraph{GUE as a free extreme value.}

We move on to describe the GUE example from the point of view of free extreme values. The CDF for the maximum is given by \eqref{maxlaw2} for GUE:
\begin{align}
\label{FrFreeGUE}
F_{H^{\lor r}}(x) = \max \left (0, r f^{(\text{GUE})}_{\text{bulk}}(x) - (r-1) \right ).
\end{align}
We use a formula $\max (0,x) = x \theta(x)$ which results in a formula equal to Eqs. \eqref{thinCDFGUE} or \eqref{FbulkGUE} obtained in the previous sections within thinning and extreme random matrix framework.

Similarly as in the thinning method, the bulk level is the only level of detail we can access through free probability. The large matrix size limit is taken implicitly so that no results related to edge-like phenomena have natural counterparts within this framework.

\section{Connections between extreme laws}
In this section, we start form  recapitulating  already observed  links as well the the new ones,  between  two  known  and two novel approaches to free (or matrix) extremes given in previous Sec. \ref{frameworks}. Then, we establish notation for classical extreme laws of Fr\'echet, Weibull and Gumbel and recall its free analogs introduced in \cite{BAV2006:FREEEXTREME} to prepare the ground for one of main results of present work -- the \emph{exponentiation} formula relating both worlds. Last section is devoted to explicit calculations of free extreme laws in models related with random matrices.

\subsection{Relating CDFs between frameworks}

Although first three frameworks of thinning method, extreme random matrices and free extreme values discussed in Sec. \ref{frameworks} start off from slightly different initial considerations, they arrive at the same CDFs given by Eqs. \eqref{Fthin}, \eqref{Frbulk} and \eqref{maxlaw2}. Extreme random matrices are equivalent only in the bulk while CDF for the free extreme values require only minor reformulation by the use of identities $\theta(af) = \theta(f) $ for $a>0$ and $\max(0,f) =f \theta(f)$. 

The last presented approach of Peak-Over-Threshold requires some work.   We present how the POT excess distribution function given by Eq. \eqref{potdef}:
\begin{align*}
    \mathcal{P}_{\text{POT}}(X < u + t | X> u) = \frac{f(u+t)-f(u)}{1-f(u)}.
\end{align*}
is related to the thinned CDF \eqref{Fthin} $\mathbf{F}_{r}(x) = r (f(x) - \alpha_r) \theta(f(x) - \alpha_r)$.

It is evident that both methods study extremes -- POT method looks at values above some threshold whereas the thinning approach focuses on a fraction $r$ of largest values drawn from the sample of $m$ observations. Thus, we relate the POT threshold $u$ to the thinning parameter $r$:
\begin{align}
\label{kurel}
f(u) = 1 - \frac{1}{r(u)},
\end{align}
via known CDF $f(x)$. This one-to-one relation dictates where one should position the threshold $u$ in order to capture a fraction $r$ of values in the sample. This relation is strict in the limit of large samples as only then the inter-sample fluctuations vanish. By the same reason, this relation makes sense for any real value of $r$. In statistics literature, the threshold value $u$ is set to be a $r$-quantile of the CDF $f$.

Now we show how with a quantile relation \eqref{kurel} between threshold $u$ and thinning size $r$, the thinned CDF given by Eq. \eqref{Fthin} has a form of the POT excess distribution function $\mathcal{P}_{\text{POT}}$ given by Eq. \eqref{potdef}. Therefore, in Eq. \eqref{Fthin} we plug $r \to r(u)$ and evaluate function at an exceedance level $x=u+t$:
\begin{align*}
\mathbf{F}_{r(u)} (u + t) = & r(u) \left (f(u+t) - 1 + \frac{1}{r(u)} \right ) \\
& \theta \left (f(u+t) - 1 + \frac{1}{r(u)} \right ).
\end{align*}
We next plug in \eqref{kurel} and find $f(u+t) - 1 + \frac{1}{r(u)} = f(u+t) - f(u)$ along with $\theta(f(u+t)-f(u)) = \theta(t)$ as $f$ is a monotonic function. Finally, we obtain
\begin{align*}
\mathbf{F}_{r(u)} (u + t) = \frac{f(t+u) - f(u)}{1-f(u)} \theta (t),
\end{align*}
which recreates the POT excess distribution function given by Eq. \eqref{potdef} with an implicit assumption that $t>0$. This expression is also exactly that of Df. 7.2 given in Ref. \cite{BAV2006:FREEEXTREME}. And so POT is equal to the remaining three approaches
\begin{align}
\label{thinPOTrel}
    \mathcal{P}_{\text{POT}}(X < u + t | X> u) = \mathbf{F}_{r(u)} (u + t)
\end{align}
through a change of variables -- instead of sample size $r$ and spectral parameter $x$ used in thinning, free extreme values or extreme random matrices we  inspect threshold $u$ with an excess parameter $t$. Parameters $r$ and $u$ are non-linearly related through CDF and Eq. \eqref{kurel}.
%

\subsection{Classic and free extreme laws}
\label{secextreme}
\paragraph{Classic extreme laws.} We first revise the classical extreme laws arising when inspecting the distribution of the largest value $y_{(1)}$ in the sample of $m$ i.i.d. variables. The thinning approach encompasses this case upon setting $n=1$ in the CDF \eqref{thinCDF2} and study the $m \to \infty$ limit:
\begin{align}
\label{classextreme}
\lim_{m\to \infty} \mathbf{F}_{m,1}(a_m + b_m x) = F^{\text{class}}(x),
\end{align}
with $m$ dependent constants $a_m$ and $b_m$ representing {\em centering} and {\em scaling} respectively. By the Fisher-Typpett-Gnedenko theorem, there exist three limiting forms of $F^{\text{max}}(x)$ depending on the properties of the CDF $f(x)$ as summarized in Tab. \ref{tab1} (see Ref. \cite{Viv2015:EVSREVIEW} for a pedagogical review). 

\begin{center}
\begin{table*}
\begin{tabular}{|c|c|c|c|}
\hline
name & Gumbel & Fr\'echet & Weibull \\
\hline
\makecell{properties of \\ PDF $p(x) = f'(x)$} & \makecell{tails falls off faster \\ than any power of $x$} & \makecell{$p(x)$ falls off as $\sim x^{-(\gamma+1)}$ \\ and is infinite} & \makecell{$p(x)$ is finite, $p(x)= 0 $ for $x> x_+$ \\ $p(x) \sim (x-x_+)^{-\gamma-1}$} \\
\hline
maximal CDF & $
F^{\text{class}}_{I}(x) = \exp \left ( -e^{-x} \right ) , x \in \mathbb{R}
$ & $
F^{\text{class}}_{II}(x) = \begin{cases}
0 & , x < 0 \\
\exp \left ( - x^{-\gamma} \right ) & , x>0
\end{cases}
$ & $F^{\text{class}}_{III}(x) = \begin{cases}
\exp \left ( -(-x)^\gamma \right ) & , x < 0 \\
1 &, x>0 
\end{cases}$ \\
\hline
$a_n$ & $f^{-1}(\alpha_n)$ & $0$ & $x_+$ \\
\hline
$b_n$ & $f^{-1}(\alpha_{ne}) - a_n$ & $f^{-1}(\alpha_n)$ & $x_+ - f^{-1}(\alpha_n)$\\
\hline
\end{tabular}
\caption{Table summarizing three classical extreme laws of Gumbel, Fr\'echet and Weibull. Functional inverse of the CDF $f$ is denoted by $f^{-1}$ and $\alpha_n = \frac{n-1}{n}$.}
\label{tab1}
\end{table*}
\end{center}


\paragraph{Free extreme laws.} Highly similar free extreme laws exist for the CDF of noncommutative (or free) random variables defined as the limit of the formula given by Eq. \eqref{maxlaw2}:
\begin{align}
\label{extrememats}
\lim_{r\to \infty } F_{H^{\lor r}} (a_r + b_r x) = F^{\text{free}}(x).
\end{align}
with some scaling and centering constants $a_r,b_r$.
The classical and free extreme laws are highly similar -- they admit the same domains of attraction, constants $a_r,b_r$ and properties of the parent distributions. The functional forms for extreme CDF's are however different and summarized in Tab. \ref{tab2}.

Since free CDF  \eqref{maxlaw2} coincides exactly with both thinning CDF \eqref{Fthin} and bulk extreme CDF \eqref{Frbulk}, the same limiting extreme laws follow. The POT method however is slightly more contrived due to change of variables \eqref{kurel} used in deriving relation \eqref{thinPOTrel}. In App. \ref{POTextreme} we work out how this variable change results in new scaling constants and the same extreme laws.

\begin{center}
\begin{table*}
\begin{tabular}{|c|c|c|c|}
\hline
name & free Gumbel & free Fr\'echet & free Weibull \\
\hline
CDF & $
F^{\text{free}}_{I}(x) = \begin{cases}
0 & , x < 0 \\
1 - e^{-x} & , x > 0
\end{cases}
$ & $
F^{\text{free}}_{II}(x) = \begin{cases}
0 &, x < 1 \\
1 - x^{-\gamma} &,  x>1
\end{cases}
$ & $ F^{\text{free}}_{III}(x) = \begin{cases}
0 & , x < -1 \\
1 -(-x)^\gamma & , x \in (-1,0) \\
1 &,  x>0
\end{cases} $ \\
\hline
$t(x)$ & 1 & $\theta(x)$ & 1 \\
\hline
$T(x)$ & $\theta(x)$ & $\theta(x-1)$ & $\theta(x+1)$ \\
\hline
examples & Free Gauss (ex. 6) & Free Cauchy $\gamma=1$ (ex. 4) & Wigner's semicircle $\gamma=3/2$ (ex. 1) \\
& & Free L\'evy-Smirnov $\gamma=1/2$ (ex. 5) & Mar\c cenko-Pastur $\gamma=3/2$ (ex. 2) \\
& & & Free Arcsine $\gamma=1/2$ (ex. 3) \\
\hline
\end{tabular}
\caption{Table summarizing free extreme laws along with concrete examples computed in Sec. \ref{examples}. Functions $t(x),T(x)$ are step functions used in the exponentiation map \eqref{exponentiation}.}
\label{tab2}
\end{table*}
\end{center}

\subsection{\textit{Exponentiation} explains relation between classical and free extreme laws}

Although classical and free (and also POT) extreme CDFs have different functional forms, they seem to be related by a striking expression
  \begin{equation}
  \begin{split}
  F^{{\rm free}}(x) & \approx 1 +\ln F^{{\rm class}(x)}, \quad \text{or}\\
  F^{{\rm class}}(x) & \approx \exp \left ( F^{{\rm free}}(x) - 1 \right ),
  \label{FvsF} 
  \end{split}
  \end{equation}
Such relation between POT and classical extreme laws has been observed in classical probability, see e.g.  \cite{EXTREMEOTHERS}. Unfortunately, it is valid only for the functional forms (see Tabs. \ref{tab1} and \ref{tab2}) and not for whole functions as their domains do not simply match up. Using the thinning method we now derive a slightly modified formula \eqref{FvsF} which corrects these domain inconsistencies. We stress that the thinning approach exemplified in formula \eqref{thinCDF} for CDF $\mathbf{F}_{m,n}$ is indispensable and encompasses both classical and free worlds. The former is attained when $n=1,m\to \infty$ while the latter as $n,m\to \infty$ with  $m/n = r$ fixed and then taking $r\to \infty$.


The classical extreme laws $F^{\text{class}}$ are found as limits of CDF \eqref{thinCDF} for $n=1$ and in the $m\to \infty$ limit:
\begin{align}
\label{lhs}
\mathbf{F}_{m,1}(x) = \left [ f(x)\right ]^m \theta (f(x)),
\end{align}
with CDF $f(x)$ and where we added step function as $f$ is a positive function. 

The free extreme laws $F^{\text{free}}$ on the other side arise from asymptotic thinned CDF given by Eq. \eqref{Fthin} where we rename $r \to m$ and write down explicitly $\alpha_m = \frac{m-1}{m}$:
\begin{align*}
\mathbf{F}_{m} (x) = m \left (f(x) - 1 + \frac{1}{m} \right ) \theta\left (f(x)-1 + \frac{1}{m} \right ),
\end{align*}
We formally solve above equation for $f(x)$:
\begin{align*}
f(x) = 1 + \frac{1}{m} \left ( \frac{\mathbf{F}_m(x)}{T_m(x)} - 1\right )
\end{align*}
and denote $T_m(x) = \theta\left (f(x)-1 + \frac{1}{m} \right )$ be the step function. We plug it back to Eq. \eqref{lhs}, set $x_m = a_m + b_m x$ with constants given in Tab. \ref{tab1} and consider the limit $m\to \infty$ to obtain the classical extreme law CDF according to Eq. \eqref{classextreme}:
\begin{align}
\label{fmax}
 & F^{{\rm class}}(x) = \lim_{m\to \infty} \mathbf{F}_{m,1}(x) = \nonumber \\
 & = \lim_{m\to \infty} \left ( \theta(f(x_m))  \left [ 1 + \frac{1}{m} \left ( \frac{\mathbf{F}_m(x_m)}{T_m(x_m)} - 1\right ) \right ]^m \right ),
\end{align}
with $x_m = a_m + b_m x$.
The ratio $\mathbf{F}_m/T_m$ in this formula is in turn expressed asymptotically by free extreme law CDF given by Eq. \eqref{extrememats} and step function $T(x)$:
\begin{align}
\lim_{m\to \infty} \frac{\mathbf{F}_m(x_m)}{T_m(x_m)} & = \frac{F^{\text{free}}(x)}{T(x)},
\end{align}
Thus, both step functions $t(x), T(x)$ admit $m$-independent form given by
\begin{equation}
\begin{split}
\label{Ttdefs}
T(x) & = \lim\limits_{m\to \infty} \theta\left (f(x_m) - \alpha_m \right ), \\
t({x}) & = \lim_{m\to \infty} \theta(f(x_m)).
\end{split}
\end{equation}
They admit three distinct forms for three extreme laws. In App. \ref{appconst} we find explicit formulas for all cases and summarize the findings in Tab. \ref{tab2} and below:
\begin{itemize}
\item $t(x) = \theta(x)$ if the parent CDF belongs to the Fr\'echet domain and $t(x)=1$ if parent CDF belongs to either Gumbel or Weibull domains whereas
\item $T(x) = \theta(x+a)$ with $a=-1$ for Fr\'echet domain, $a=0$ for Gumbel domain and $a=1$ for Weibull domain.
\end{itemize} 
After plugging these definitions back into Eq. \eqref{fmax}, we use finally the exponentiation formula $\lim\limits_{m\to \infty} (1+x/m)^m = e^x$ and the corrected formula \eqref{FvsF} relating free and classical extreme laws reads
\begin{align}
\label{exponentiation}
F^{{\rm class}}(x) = t(x) \exp \left (  \frac{F^{\text{free}}(x)}{T(x)} - 1 \right ),
\end{align}
where step functions $t$ and $T$ are given in Tab. \ref{tab2}. 

Presence of step functions in the denominator is a formal notation which becomes evident by rewriting free CDFs of Tab. \ref{tab2} with the use of step functions:
\begin{itemize}
\item Gumbel domain gives $F^{\text{free}}_{I}(x) = \theta(x) (1-e^{-x})$,
\item Fr\'echet domain gives $F^{\text{free}}_{II}(x) = \theta(x-1) (1-x^{-\gamma})$,
\item Weibull domain gives $F^{\text{free}}_{III}(x) = \theta(x+1) \left [ 1 - \theta(-x)(-x)^\gamma\right ]$.
\end{itemize}
and realizing how their pure functional forms can be expressed as a ratio.

\paragraph{POT extreme laws.} Finally, the POT formalism is in the same region as free extreme laws and so formally all formulas shown in this section hold also for POT approach. Details concerning different scaling constants were given in App. \ref{POTextreme}.

\subsubsection{Examples}
\label{examples}

Finally, due to the operational simplicity of the thinning theorem, we are able to give several explicit examples of free extreme laws following from spectral densities of large random matrix models. Majority  of well-known models is defined on finite spectral support, alike Wigner's semicircle or Mar\c{c}enko-Pastur distribution. Via exponentiation argument, they belong therefore to free Weibull class, which we show  by explicit calculation. The free Fr\'{e}chet class is more subtle, since the spectral density has to vanish as a  power law and the support is not limited. The so-called Bercovici-Pata construction \cite{BERCOVICIPATA}, being the analogue of L\'{e}vy heavy-tailed distributions in classical probability, provides explicit examples. We consider two exotic random matrix models, corresponding to free Cauchy and free L\'{e}vy-Smirnov distribution, and by explicit calculation we show that they realise  extreme statistics of the  free Fr\'{e}chet class. 
The last class, free Gumbel distribution, turned out to be most demanding to find, despite being relatively common in classical probability, as realised e.g. by Gaussian or Poisson distributions. Here as an example we used recent works \cite{BOZEJKOOLD,Tie2001:FREEGAUSSMAT}.
This last example shows that the standard folklore of calling Wigner semicircle as a "free Gaussian" has to be used with care. 

In what follows we use the extreme CDF found in all discussed frameworks:
\begin{align*}
\mathbf{F}_r(x) = r \left (f(x) - \alpha \right ) \theta\left (f(x) - \alpha \right ).
\end{align*} 
and moreover we introduce the corresponding PDF:
\begin{align*}
\mathbf{p}_r(x) = \frac{d}{dx} \mathbf{F}_r(x) = r \rho(x) \theta\left (f(x) - \alpha \right ),
\end{align*}
with density $\rho(x) = f'(x)$.

\paragraph{ Wigner's semicircle law (free Weibull domain).}
An example of GUE discussed in Sec. \ref{GUE} belongs to the free Weibull domain. To show this, we choose scaling parameters $a_r = 2, b_r = a_r - f^{-1}(\alpha)$ with CDF given by Eq. \eqref{fGUEbulk}, set $x = a_r + b_r \tilde{x}$ and find $\theta(f(x) - \alpha) = \theta (x-f^{-1}(\alpha)) =  \theta\left ((2-f^{-1}(\alpha))(\tilde{x}+1) \right ) = \theta(\tilde{x}+1)$. With the approximation $f^{-1}(\alpha) \sim 2 - \left ( \frac{3\pi}{2k} \right ) ^{\frac{2}{3}}$ we find the extreme PDF:
\begin{align*}
\lim_{r\to \infty} \mathbf{p}_r(a_r + b_r \tilde{x}) b_r d\tilde{x} =  \frac{3}{2} (-\tilde{x})^{\frac{1}{2}} \theta(\tilde{x}+1) \theta(-\tilde{x}) d\tilde{x},
\end{align*}
where the second Heaviside theta function arises by truncating the semicircle law as $\theta(2-x) = \theta(-\tilde{x})$. The extreme CDF therefore reads:
\begin{align*}
F_{III}^{\text{GUE}}(x) = \begin{cases}
0 & , \qquad x < -1 \\
1 -(-x)^{3/2} & , \qquad x \in (-1,0) \\
1 &, \qquad x>0
\end{cases},
\end{align*}
and is an example of the free Weibull distribution of Tab. \ref{tab2} with parameter $\gamma = 3/2$. 
\paragraph{Mar\c cenko-Pastur law (free Weibull domain).}
We check that Mar\c cenko-Pastur law for Wishart matrices also belongs to the free Weibull domain. To this end, we write down both parent PDF and CDF for rectangularity $0<s<1$:
\begin{align*}
\rho_{\text{MP}}(x) = & \frac{1}{2\pi s x} h(x), \qquad x_\pm = (1\pm \sqrt{s})^2, \\
f_{\text{MP}}(x) = & \frac{1}{2} + \frac{h(x)}{2\pi r} - \frac{1+s}{2\pi s} \arctan \left ( \frac{1+s-x}{h(x)} \right ) + \\
& + \frac{1-s}{2\pi s} \arctan \left ( \frac{(1-s)^2 - x(1+s)}{(1-s) h(x)} \right ),
\end{align*}
with $h(x) = \sqrt{(x_+ - x)(x-x_-)}$. We expand CDF around the endpoint $f_{\text{MP}}(x_+ - \delta) \sim 1 - \frac{2}{3\pi} \frac{1}{(1+\sqrt{s})^2s^{3/4}} \delta^{3/2}$ and the inverse CDF around unity $f^{-1}_{\text{MP}}\left (1-\frac{1}{r} \right ) \sim x_+ - \delta$ to find an
approximate formula to the latter
\begin{align*}
f^{-1}_{\text{MP}} (\alpha) \sim x_+ - \left ( \frac{3\pi}{2r} \right )^{2/3} (1+\sqrt{s})^{4/3} \sqrt{s}.
\end{align*}
We compute the rescaled PDF for scaling parameters $a_r = x_+, b_r = x_+ - f^{-1}_{\text{MP}}(\alpha)$ and find that it belongs to the free Weibull domain with parameter $\gamma=3/2$:
\begin{align*}
F_{III}^{\text{MP}}(x) = \begin{cases}
0 & , \qquad x < -1 \\
1 -(-x)^{3/2} & , \qquad x \in (-1,0) \\
1 &, \qquad x>0
\end{cases}.
\end{align*}
\paragraph{Free Arcsine (free Weibull distribution).} Arcsine distribution is given by the PDF $\rho(x)=\frac{1}{\pi} \frac{1}{\sqrt{x(1-x)}}$. In the classical probability, this is a  special case of the  beta distribution $\beta \left (\frac{1}{2}, \frac{1}{2} \right )$. In the free probability, such spectral measure corresponds to the free convolution of the identical, mutually free  discrete measures concentrated on two points (Dirac deltas), i.e. $\frac{1}{2} ( \delta(x) +\delta(x-1/2))$~(see Ref. \cite{DKV}).  A rerun of arguments  presented for semicircle (with 
 $f_{{\rm arcsine}}(x)= \frac{2}{\pi} \arcsin \sqrt{x}$), yields immediately to the conclusion, that extreme statistics for arcsine law  belongs to the Weibull domains (both classical and free) with $\gamma=1/2$.

\paragraph{Free Cauchy (free Fr\'{e}chet domain).}
In free probability, there exists the whole class of spectral distributions, which are stable under the free convolution, modulo the affine transformation. They form exactly the analogue of L\'{e}vy  heavy (fat) tail distributions in classical probability theory. This one-to-one analogy is called Bercovici-Pata bijection~\cite{BERCOVICIPATA}. As the simplest example in the free probability context, the following PDF and CDF are considered:
\begin{align*}
\rho_C(x) & = \frac{1}{\pi} \frac{1}{1+x^2}, \\
f_C(x) & = \frac{1}{2} + \frac{1}{\pi} \arctan(x), \qquad f^{-1}_C(x) = -\cot (x \pi).
\end{align*}
This is the symmetric, spectral Cauchy distribution.  The realization of such free heavy-tailed ensembles is non-trivial, e.g. the potential, which by the entropic argument yields Cauchy spectrum, reads explicitly~\cite{FREELEVY}
\begin{align*}
V(\lambda)= \frac{1}{2} \ln (\lambda^2 +1)
\end{align*}
so it is non-polynomial - note, that for the Gaussian ensembles $V(\lambda) \sim \lambda^2$.
However, to get the extreme law we do not need at any moment the form of the potential. According to Tab. \ref{tab1}, we choose $a_r = 0, b_r = f^{-1}_C(\alpha)$, compute $\theta(a_r + b_r \tilde{x} - f_C^{-1}(\alpha) ) = \theta (\tilde{x} - 1) $ and find the PDF:
\begin{align*}
\lim_{r\to \infty} \mathbf{p}_r(a_r + b_r \tilde{x}) b_r d\tilde{x} = \frac{1}{\tilde{x}^2} \theta (\tilde{x} - 1) d\tilde{x}.
\end{align*}
Upon integration the extreme CDF in turn reads:
\begin{align*}
F_{II}^{\text{C}}(x) = \begin{cases}
0 & , \qquad x < 1 \\
1- x^{-1} &, \qquad x>1
\end{cases}
\end{align*}
and belongs to the free Fr\'echet class of Tab. \ref{tab2} with $\gamma = 1$. 
\paragraph{L\'evy-Smirnov distribution (free Fr\'{e}chet domain).} Free L\'{e}vy-Smirnov distribution is another, simple example belonging to the free Fr\'{e}chet domain. It is realized through an entropic argument with a confining potential $V(\lambda)=\frac{1}{2\lambda} +\ln \lambda$~\cite{FREELEVY}. The PDF and CDF read respectively:
\begin{align*}
\rho_{\text{LS}}(\lambda) & = \frac{1}{2 \pi} \frac{\sqrt{4\lambda-1}}{\lambda^2}, \qquad \lambda \in \left ( \frac{1}{4}, \infty \right ), \\
f_{\text{LS}}(x) & = \frac{2}{\pi} \arccos \left ( \frac{1}{2\sqrt{x}} \right ) - \frac{\sqrt{4x-1}}{2\pi x}.
\end{align*}
The inverse CDF is found by expanding CDF $f_{\text{LS}}(x) \sim 1 - \frac{2}{\pi} \sqrt{\frac{1}{x}}$ around $x\to \infty$ which results in
\begin{align*}
f^{-1}_{\text{LS}}(\alpha) & \sim \left ( \frac{2r}{\pi} \right )^2.
\end{align*}
The scaling parameters read $a_r = 0$, $b_r = f^{-1}_{\text{LS}}(\alpha)$, the limiting extreme CDF reads
\begin{align*}
F_{II}^{\text{LS}}(x) = \begin{cases}
0 & , \qquad x < 1 \\
1- x^{-\frac{1}{2}} &, \qquad x>1
\end{cases}
\end{align*}
and belongs to the free Fr\'echet class with $\gamma = 1/2$.
\begin{center}
\begin{figure}
\includegraphics[scale=.7]{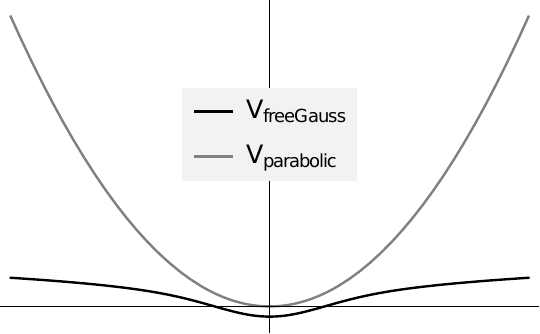}
\caption{Plot showing weak confinement property of free Gaussian potential \eqref{freegauss} in comparison   to the Gaussian Unitary Ensemble parabolic shape $V_{\text{{\rm{ parabolic}}}} = \lambda^2/2$. The former results in bell-shaped spectral density with infinite support while the latter is a prime example of semicircular spectral density with finite support.}
\label{potentialfig}
\end{figure}
\end{center}
\paragraph{Free Gaussian (free Gumbel domain).}
To apply our procedure for this case, we have to choose the spectral distribution whose tails fall faster  than any power of $x$. We can use the powerful result~\cite{BOZEJKOOLD,BOZEJKO}, noticing that the normal distribution is freely infinitely divisible.  This implies, that there exists  a random $ N \times N$ matrix ensemble, whose spectrum in the large $N$ limit approaches the normal distribution.   
Entropic argument  can even help to find the  shape of the confining potential yielding such distribution~\cite{Tie2001:FREEGAUSSMAT}
\begin{align}
\label{freegauss}
V(\lambda) = c + \frac{\lambda^2}{2} ~ {}_2F_2 \left (1,1;\frac{3}{2},2; -\frac{\lambda^2}{2} \right ),
\end{align}
where $c = - \frac{\gamma + \log 2}{2}$ and the potential is a solution to $V(\lambda) =  \frac{1}{\sqrt{2\pi}}\int_{-\infty}^\infty e^{-x^2/2} \ln |x-\lambda| dx $. In Fig. \ref{potentialfig} we plot potential $V(\lambda)$ against a parabolic function showing its weakly confining property. It is amusing to note, that the Green's function corresponding to free Gaussian is equal,  modulo the sign, to the famous and well-studied plasma dispersion function $Z$~\cite{ZFUNCTION},  which allows e.g. to study the subtle asymptotics of  the resolvent. 
Luckily,  in our thinning model we do not need the shape of the potential to find the free extreme laws. 
The resulting PDF, CDF, and inverse CDF (quantile) for the spectral normal distribution read, respectively:
\begin{align*}
\rho_{\text{G}}(x) & = \frac{1}{\sqrt{2\pi}} e^{-x^2/2}, \\
f_{\text{G}}(x) & = \frac{1}{2} \left (1+\text{erf} (x/\sqrt{2}) \right ), \quad f^{-1}_{\text{G}}(x) = \sqrt{2} \text{erf}^{-1} (2x-1).
\end{align*}
According to the Table \ref{tab1}, we set $a_r = f^{-1}_{\text{G}}(1-1/r), b_r = f^{-1}_{\text{G}}(1-1/(er)) - f^{-1}_{\text{G}}(1-1/r)$ and so with $x = a_r + b_r \tilde{x}$ we have to perform the limit 
\begin{align*}
\lim_{r\to \infty} \mathbf{p}_r(a_r + b_r \tilde{x}) b_r d\tilde{x} =  \lim_{r \rightarrow \infty} \frac{r}{\sqrt{2\pi}} b_r e^{-[a_r +b_r \tilde{x}]^2/2}d\tilde{x}.
\end{align*}
The limit is subtle, since the inverse error function develops the singularity when its argument approaches unity
\begin{align*}
\text{erf}^{-1}(z)|_{z \rightarrow 1} \sim \frac{1}{\sqrt{2}} \sqrt{ \ln \left( g(z) \right) -\ln \left (\ln  \left(g(z)\right) \right )}.
\end{align*} 
with $g(z) = \frac{2}{\pi(z-1)^2} $. We set $r=\sqrt{2\pi} e^{u/2}$ and find asymptotic series for both scaling parameters $a_r \sim \sqrt{u- \ln u}$ and $b_r \sim \sqrt{2+u-\ln(2+u)}-\sqrt{u-\ln u}$. These asymptotic expansions result in $a_r^2 \sim  u-\ln u, a_r b_r \sim 1, b_r^2  \sim \frac{1}{u}$ and $ \ln b_r \sim -\frac{1}{2} \ln u$ which makes all divergent terms cancel out and  only the $a_r b_r \sim 1$ survives, yielding 
\begin{align*}
\lim_{r \rightarrow \infty} \frac{r}{\sqrt{2\pi}} b_r e^{-[a_r +b_r \tilde{x}]^2/2}d\tilde{x}= e^{-\tilde{x}}\theta ({\tilde{x}}) d \tilde{x},
\end{align*} 
which in turn gives the CDF:
\begin{align*}
F_{I}^{\text{G}}(x) = \begin{cases}
0 & , \qquad x < 0 \\
1- e^{-x} &, \qquad x>0
\end{cases},
\end{align*}
an instance of free Gumbel domain of Tab. \ref{tab2}.

 Other examples include e.g. several free infinite divisible gamma  distributions~\cite{HASEBE},  with the simplest $\rho(x)=e^{-x}$ for non-negative $x$. Since  $f(x)=1-e^{-x}$, the application of scaling and centering formulae from Table~I yields, $a_r=\ln r$ and $b_r=1$,  which  trivially reproduces  the free Gumbel CDF. 
 


We stress that the same functional form of the PDF may lead to either classical or free extreme law, depending if the PDF represents the one-dimensional, classical probability or represents spectral PDF of the ensemble of asymptotically large matrices. Examples~3, 4 and 6 show it explicitly, for each domain: Weibull, Fr\'{e}chet and  Gumbel, respectively. 

\section{Conclusions and outlook}

In this work on extreme matrices we have devised a new \emph{thinning} method which, in contrast with approaches introduced previously, is able to bridge the gap between classical extreme values and free (or matrix) extreme values. Through this link we establish an explicit exponentiation map between classical extreme laws of Weibull, Fr\'echet and Gumbel and free Weibull, free  Fr\'echet and free Gumbel. On top of that, we show that also Peak-Over-Threshold method is related to thinning approach through a simple change of variables. Finally, we provide an approach of extreme random matrices which in turn enables refined questions about the spectra of extreme matrices. Thinning method provides an operational language which we elucidate by showing several explicit examples of extreme laws of random matrix inspired models.

Studies of extreme matrices started relatively recently and so many questions remain unanswered. Among the most promising is a free analog of Tracy-Widom law or free Airy-type behaviour near the spectral edge of the extreme matrix.



\section{Acknowledgments}
The authors appreciate discussions  with M. Bo\.{z}ejko, S. Majumdar and D. V. Voiculescu, and comments from P. Warcho\l{} and W. Tarnowski. They also thank Ilan Roth for pointing the connection of the Green's function for free Gaussian to the famous plasma dispersion function $Z$.   The research was supported by the MAESTRO DEC-2011/02/A/ST1/00119 grant of the Polish National Center of Science (2017-2018) and  by the TEAMNET  POIR.04.04.00-00-14DE/18-00 grant of the Foundation for Polish Science (2019-2020).


\appendix
\section{Derivation of Eq. \eqref{Fr}}
\label{FrApp}
We start off from the definition \eqref{Frdef} and plug in the formula \eqref{genp2}:
\begin{align}
\label{Frstart}
F_{N,r}(x) = \frac{1}{N} \sum_{k=1}^N \sum_{i=0}^{k-1} T_i,
\end{align}
with
\begin{align*}
T_i = \sum_{ \sigma,\delta } [x_{\sigma(1)} ... x_{\sigma(i)} ]_J [x_{\delta(1)} ... x_{\delta(rN-i)} ]_{\bar{J}} \mathbf{P}(x_1,...,x_{rN}),
\end{align*}
where we introduced a succinct integral notation where $ [ y_1 ... y_k ]_J = \int_x^{\infty} dy_1 \cdots \int_x^{\infty} dy_k $ and $\bar{J} = \mathbb{R} - J$. 
Firstly, we reintroduce the notation highlighting the initial $r$ matrix ensembles and factor both jPDF and integrals:
\begin{align*}
& \left [x_{\sigma(1)} ... x_{\sigma(i)} \right ]_J \left [x_{\delta(1)} ... x_{\delta(rN-i)} \right ]_{\bar{J}} \mathbf{P}(x_1,...,x_{rN}) = \\
= & \prod_{k=1}^r \left [\lambda^{(k)}_{\sigma_k(1)} ... \lambda^{(k)}_{\sigma_k(j_k)} \right ]_J \left [\lambda^{(k)}_{\delta_k(1)} ... \lambda^{(k)}_{\delta_k(N-j_k)}  \right ]_{\bar{J}} \\
& P^{(k)}_N\left (\lambda_1^{(k)}...\lambda_N^{(k)} \right ),
\end{align*}
where $\sigma_k$ and $\delta_k$ refer to subsets of $\sigma$ and $\delta$ respectively which correspond to the $k$-th set of $\lambda$'s as dictated by the dictionary \eqref{trans}. Furthermore, the summation over combinations $\sigma,\delta$ is likewise split by the aforementioned dictionary :
\begin{align*}
\sum_{\sigma,\delta} = \sum_{\substack{\{\sigma_1,\delta_1\},...,\{\sigma_r,\delta_r\}\\ j_1 + ... + j_r = i}},
\end{align*}
where the only constraint is the conservation of the sizes of the set $\sigma$ and all subsets $\sigma_1...\sigma_r$. We can now factorize $T_i$ almost completely:
\begin{align*}
T_i = & \sum_{\sigma,\delta} \prod_{k=1}^r \left [\lambda^{(k)}_{\sigma_k(1)} ... \lambda^{(k)}_{\sigma_k(j_k)} \right ]_J \left [\lambda^{(k)}_{\delta_k(1)} ... \lambda^{(k)}_{\delta_k(N-j_k)}  \right ]_{\bar{J}} P_N^{(k)},
\end{align*}
with $P_N^{(k)} = P_N^{(k)}\left (\lambda_1^{(k)}...\lambda_N^{(k)} \right )$. We readily see that all $r$ terms in the sum are coupled only through the constraint $j_1+...+j_r = i$. Lastly, we rewrite $k$-th sum by using the symmetry of the jPDF and introduce a combinatorial factor $\binom{N}{j_k}$:
\begin{align*}
& \sum_{\{\sigma_k,\delta_k\} }\left [\lambda^{(k)}_{\sigma_k(1)} ... \lambda^{(k)}_{\sigma_k(j_k)} \right ]_J \left [\lambda^{(k)}_{\delta_k(1)} ... \lambda^{(k)}_{\delta_k(N-j_k)}  \right ]_{\bar{J}} P_N^{(k)} = \\ & = \sum_{j_k=0}^i \binom{N}{j_k} \left [\lambda^{(k)}_{1} ... \lambda^{(k)}_{j_k} \right ]_J \left [\lambda^{(k)}_{j_k+1} ... \lambda^{(k)}_{N}  \right ]_{\bar{J}} P_N^{(k)} = \\
& = \sum_{j_k=0}^i E_N(j_k;x),
\end{align*}
where the term $E_N(j_k;x)$ is the gap function introduced in Eq. \eqref{En}. Thus, the term $T_i$ reads
\begin{align*}
T_i = \sum_{\substack{j_1...j_r = 0 \\ j_1+...+j_r = i}}^i \prod_{k=1}^r E_N(j_k;x).
\end{align*}
We can now plug in above equation into Eq. \eqref{Frstart} and use a simple identity:
\begin{align}
\label{ident0}
\sum_{k=1}^N \sum_{i=0}^{k-1} S_i = \sum_{n=0}^{N-1} (N-n)S_n,
\end{align}
to finally arrive at the Eq. \eqref{Fr}
\begin{align*}
F_{N,r}(x) = \frac{1}{N} \sum_{n=0}^{N-1} (N-n) \sum_{\substack{j_1...j_r = 0\\ j_1+...+j_r = n}}^n \prod_{k=1}^r E_N(j_k;x). 
\end{align*}

\section{Derivation of Eq. \eqref{F1form}}
\label{F1App}
To obtain the formula \eqref{F1form} in a straightforwad manner, we use the following relation between gap functions and correlation functions:
\begin{align}
\label{ident}
& \sum_{m=0}^{N-n} \frac{(m+n)!}{m!} E_N(m+n;x) = \nonumber \\
& = \int_{x}^{\infty} dx_1 ... \int_{x}^{\infty} dx_n \rho^{(n)}_N (x_1,...,x_n),
\end{align}
where $\rho_N^{(n)} (x_1...x_n) = \det K_N(x_i,x_j)_{i,j=1...n}$ are the $n$-point correlation functions. We split $F_{N,1}$ into two parts:
\begin{align*}
F_{N,1}(x) = \frac{1}{N} \sum_{n=0}^{N-1} (N-n) E_N(n;x) = f_1 + f_2, 
\end{align*}
with $f_1 = \sum_{n=0}^{N-1} E_N(n;x)$ and $f_2 = - \frac{1}{N} \sum_{n=0}^{N-1} n E_N(n;x)$. The term $f_1$ is found by setting $n=0$ in identity \eqref{ident}:
\begin{align*}
\sum_{m=0}^{N} E_N(m;x) = 1 \quad \to \quad f_1 = 1 - E_N(N;x),
\end{align*} 
and the second $f_2$ by plugging $n=1$ to Eq. \eqref{ident}:
\begin{align*}
\sum_{m=1}^N m E_N(m;x) = \int_{x}^{\infty} dy \rho_N^{(1)}(y)
\end{align*}
giving
\begin{align*}
f_2 = -\frac{1}{N} \int_{x}^{\infty} dy \rho_N^{(1)}(y) + E_N(N;x).
\end{align*}
We add both contributions and use the integral $\int_{-\infty}^\infty \rho_N^{(1)}(y) dy = N$ to finally arrive at Eq. \eqref{F1form}:
\begin{align*}
F_{N,1}(x) = 1 - \frac{1}{N} \int_{x}^{\infty} dy \rho_N^{(1)}(y) = \frac{1}{N} \int^{x}_{-\infty} dy \rho_N^{(1)}(y).
\end{align*}

\section{Gap functions $E_N(j;x)$ in different regimes}
\label{gapasymptotics}
To advance our discussion for general $r$, we first address behaviour of gap functions $E_N(j;x)$ in greater detail. We inspect the bulk and edge regimes as two typical examples of two different yet universal spectral behaviours. We base our discussion on an exact formula relating gap functions to $k$-point correlation functions $\rho_N^{(k)}$:
\begin{align}
\label{Encorr}
E_N(k;x) = \sum_{l=k}^{N} \frac{(-1)^{l-k}}{k!(l-k)!} \int\limits_x^\infty d\lambda_1 ... d\lambda_{l} \rho_N^{(l)}(\lambda_1...\lambda_{l} ).
\end{align}
\paragraph{Gap function in the bulk.}
We assume the matrix model parameters are re-scaled so that the spectral domain is bounded as $N \to \infty$. Then the bulk regime is probed when $x \in \mathcal{O}(1)$. On this scale, the $k$-point correlation functions decouple and their dominant contribution comes from 1-point correlations:
\begin{align*}
\rho^{(l)}_N (x_1...x_l) \sim N^l \rho_{\text{bulk}}^{(1)}(x_1) \cdots \rho_{\text{bulk}}^{(1)}(x_l). 
\end{align*}
Consequently, we combine this approximation with the formula \eqref{Encorr} so that $E_N(k;x) \sim E_{\text{bulk}}(k;x)$ with:
\begin{align}
\label{Enbulk}
E_{\text{bulk}}(k;x) = e^{-N(1-f(x))} \frac{[N(1-f(x))]^k}{k!},
\end{align}
valid for $x \in \mathcal{O}(1)$ and where the spectral CDF $f$ is given by:
\begin{align*}
f(x) = \frac{1}{N} \int_{-\infty}^x \rho_{\text{bulk}}^{(1)}(y) dy.
\end{align*}
We stress that in the bulk, all eigenvalue correlations are absorbed into the spectral density alone and its CDF $f(x)$. This is reflected in result \eqref{Enbulk} having a form of PDF of an inhomogenous Poisson process with variable intensity function $\rho_{\text{bulk}}^{(1)}$.

\paragraph{Gap function near the (soft) edge.}
Typically the bulk regime heralds its limitations by existence of a (soft) edge as the point where the bulk spectral density vanishes. We zoom in the vicinity of such points by setting $x = x_{\text{edge}} + \sigma N^{-\alpha}$ with an appropriate scaling exponent $\alpha$. In this regime however, all correlations contribute and so $ E_{\text{edge}}(k;\sigma) = \lim_{N\to \infty}  E_N(k;x_{\text{edge}} + \sigma N^{-\alpha})$ is a sum of all correlations:
\begin{align}
\label{Enedge}
E_{\text{edge}}(k;\sigma) = \frac{1}{k!} \sum_{l=0}^{\infty} \frac{1}{l!} (-1)^{l} C_{l+k}(\sigma), \qquad \sigma \in \mathbb{R},
\end{align}
where the correlation coefficients $C_l$ are given in terms of an edge kernel $k_{\text{edge}}$:
\begin{align*}
C_{l}(\sigma) = \int_\sigma^{\infty} dx_1 ... dx_l \det k_{\text{edge}}(x_i,x_j)_{i,j=1...l}.
\end{align*}

\section{Asymptotic form of Eq. \eqref{Frbulk}}
\label{Frbulkder}
We calculate asymptotic form of the following sum
\begin{align*}
F_{r}(x) \sim e^{-Nr(1-f)} \frac{1}{N} \sum_{n=0}^{N-1} (N-n) \frac{1}{n!} \left [ Nr(1-f) \right ]^{n},
\end{align*}
where $r>1$ and $f\in(0,1)$. We separate the trivial prefactor from $F_r =  e^{-Nr(1-f)} \frac{1}{N} S$ and study the sum:
\begin{align*}
S = \sum_{n=0}^{N-1} (N-n) \frac{1}{n!} \left [ Nr(1-f) \right ]^{n}.
\end{align*}
Next we apply the Euler-Maclaurin formula:
\begin{align*}
S \sim N^2 \int_0^1 dy \frac{1-y}{\Gamma(Ny+1)} \left ( Nr(1-f) \right )^{Ny} ,
\end{align*}
and cast into a form
\begin{align*}
S \sim N^2 \int_0^1 dy e^{Nf(y)} g(y),
\end{align*}
where we expanded Gamma function and found $f(y) = y\ln (r(1-f)) +y - y \ln y $, $g(y) = \frac{1-y}{\sqrt{2\pi N y}} $. From equation $f'(y)=0$ we compute the saddle point $y_*=r(1-f)$ which is always positive $y_*>0$ however not always lies inside the integration interval $(0,1)$. Specifically, $y_*<1$ if $ f > \alpha$ with $\alpha = \frac{r-1}{r}$ and $y_*>1$ when $f<\alpha$. We compute the integral around $y = y_* + \delta/\sqrt{N}$:
\begin{align*}
S_{f>\alpha} \sim  N e^{Nr(1-f)} (1-r(1-f)).
\end{align*}
Likewise, when $f<\alpha$ the main contribution comes from the $y=1$ endpoint:
\begin{align*}
S_{f<\alpha} \sim N^2 e^{Nf(1)} g(1) = 0.
\end{align*}
The integral is therefore given by
\begin{align*}
S \sim  N e^{Nr(1-f)} (1-r(1-f))\theta(f-\alpha).
\end{align*}
We collect all coefficients and go back to $F_r$:
\begin{align*}
F_r \sim r (f-\alpha) \theta(f-\alpha).
\end{align*}

\section{Asymptotic form of Eq. \eqref{Frhoedge}}
\label{Frhoedgeder}
We start off from
\begin{align*}
S_\rho = \frac{1}{N} \sum_{n=0}^{N-1} (N-n) (1-f)^{n} f^{N\rho-n}  \binom{N\rho}{n},
\end{align*}
where we denote $f = E_{\text{edge}}(0;\sigma)$.

For $\rho>1$ and $f\in(0,1)$ we approximate the sum in $S_\rho = \frac{1}{N} S$:
\begin{align*}
S = \sum_{n=0}^{N-1} (N-n) (1-f)^{n} f^{N\rho-n}  \binom{N\rho}{n}.
\end{align*}
We put Gamma functions:
\begin{align*}
S = f^{N\rho} (N\rho)! \sum_{n=0}^{N-1} \frac{N-n}{\Gamma(N\rho-n+1)\Gamma(n+1)} \left ( \frac{1-f}{f} \right )^n,
\end{align*}
and use Euler-Maclaurin formula:
\begin{align*}
S \sim f^{N\rho} (N\rho)! N^2 \int_0^1 dy \frac{(1-y)(1-f)^{Ny} f^{-Ny}}{\Gamma(N(\rho-y)+1)\Gamma(Ny+1)}.
\end{align*}
We set $S = f^{N\rho} (N\rho)! N^2 I$ and the integral $I$ is given by:
\begin{align*}
I = \int_0^1 dy \frac{1-y}{\Gamma(N(\rho-y)+1)\Gamma(Ny+1)} \left ( \frac{1-f}{f} \right )^{Ny},
\end{align*}
Integral is in the form $I \sim \int_0^1 dy e^{Nf(y)} g(y)$ where we expanded both Gamma functions and found $f(y) = y\ln \frac{1-f}{f} - (\rho-y) \ln (\rho-y) -y \ln y$, $g(y) = \frac{e^{N\rho}}{2\pi N^{N\rho+1}} \frac{1-y}{\sqrt{(\rho-y)y}}$. From equation $f'(y)=0$ we compute the saddle point $y_*=\rho(1-f)$ which is always positive $y_*>0$ however not always lies inside the integration interval $(0,1)$. Specifically, $y_*<1$ if $ f > \alpha_\rho$ with $\alpha = \frac{\rho-1}{\rho}$ and $y_*>1$ when $f<\alpha_\rho$. We compute the integral around $y = y_* + \delta/\sqrt{N}$:
\begin{align*}
I_{f>\alpha_\rho} \sim \frac{1}{N} \frac{e^{N\rho}(f\rho)^{-N\rho}}{N^{N\rho}} \frac{1-\rho(1-f)}{\sqrt{2 N \pi \rho} }.
\end{align*}
Likewise, when $f<\alpha$ the main contribution comes from the endpoint $y=1$:
\begin{align*}
I_{f<\alpha_\rho} \sim e^{Nf(1)} g(1) = 0.
\end{align*}
The integral is therefore given by
\begin{align*}
I \sim \frac{1}{N} \frac{e^{N\rho}(f\rho)^{-N\rho}}{N^{N\rho}} \frac{1-\rho(1-f)}{\sqrt{2 N \pi \rho} }\theta(f-\alpha_\rho).
\end{align*}
Lastly we collect coefficients and go back to $S_\rho$:
\begin{align*}
S_\rho \sim \rho (f-\alpha_\rho) \theta(f-\alpha_\rho). 
\end{align*}

\section{Primer on free extreme values}
\label{appe}
Free extreme values were introduced in general operator language in Refs. \cite{BAV2006:FREEEXTREME,BAK2010:FREEEXTREME2} and the special case of extreme matrices were discussed in detail in Ref. \cite{BG2010:MATRIXEXTREME}. We review these findings in what follows. 

To quantify extremal features in a set of observables, a notion of order is indispensable -- a characteristic of being either largest or smallest is needed. Although a natural inequality operator exists for real numbers (giving rise to standard extreme value statistics) for other objects it may not be the case. Already the complex plane lacks such an ordering; $z_1 \geq z_2$ for two complex numbers does not bear any natural interpretation. In particular, defining the order either by modulus or by comparing real and imaginary parts separately does not produce satisfactory results. To assess such problems in full generality, the \emph{order theory} developed two notions of \emph{partially} and \emph{totally} ordered set. A \emph{partially ordered set} is endowed with a binary relation $\leq$ satisfying three conditions satisfied by all elements $a,b,c$ in the set:
\begin{itemize}
\item $a \leq a$ (reflexivity)
\item $a \leq b$ and $b \leq a$ $\rightarrow$ $a = b$ (antisymmetry)
\item $a \leq b$ and $b \leq c$ $\rightarrow$ $a \leq c$ (transitivity)
\end{itemize}
A \emph{totally ordered set} is a partially ordered set with an additional property -- for each $a,b$
\begin{itemize}
\item either $a \leq b$ or $b\leq a$ (totality)
\end{itemize}

Given these definitions, we reconsider the example of complex numbers -- modulus-based ordering does not satisfy the antisymmetry property and the real-and-imaginary-based ordering is only partial. We are now ready to define extreme cases of more exotic objects like matrices.

We focus on extremes in the space of Hermitian matrices. Consider finite $N\times N$ matrices with $N$ distinct eigenvalues $\lambda_i$ and eigenvectors $\ket{\psi_i}$. A spectral order is defined for Hermitian matrices $H_a,H_b$ in terms of the spectral projections:
\begin{align*}
H_a \prec H_b \quad \longleftrightarrow \quad E(H_a;[t,\infty)) \leq E(H_b;[t,\infty)),
\end{align*}
where the spectral projection for a matrix $H_a$ is given by $E(H_a;[t,\infty)) = \sum_{i=1}^N \ket{\psi_i} \bra{\psi_i} \theta(\lambda_i - t)$ and $A \leq B$ operation on the space of Hermitian projections is defined as $\forall{x}:~ x^T (B-A) x \geq 0$. With this ordering, the definitions of min ($\land$) is $E (H_a \land H_b;[t,\infty)) = E (H_a;[t,\infty)) \land E (H_b;[t,\infty))$ and for max ($\lor$) is $E (H_a \lor H_b;[t,\infty)) = E (H_a;[t,\infty)) \lor E (H_b;[t,\infty))$.

\section{POT extreme laws}
\label{POTextreme}
In this appendix we show how the Peak-Over-Threshold method produces a highly similar family of extreme law under the name of generalized Pareto distribution (see Ref. \cite{BH1974:POTLAWS}). By definition \eqref{potdef}, the limiting distribution reads:
\begin{align*}
\lim_{u \to \infty} \mathcal{P}_{\text{POT}}(X < u + \tilde{a}_u + \tilde{b}_u x | X> u) = F^{\text{POT}}(x),
\end{align*}
for some constants $\tilde{a}_u, \tilde{b}_u$ which although different than in the classical and free cases, we can fix them by demanding that extreme-matrix and POT limiting distributions are exactly the same $F^{\text{POT}}(x) = F^{\text{free}}(x)$. This is possible as the constants are not unique. Moreover, from the point of view of previously derived equivalencies this is hardly a surprise - both free and POT extreme laws ought to be related as both are expressed by the same extremal CDF. 

Relation between the constants are found by
\begin{align*}
F^{\text{free}}(x) =  \lim_{r\to \infty}  \mathbf{F}_r (a_r + b_r x) = \lim_{u \to \infty} \mathbf{F}_{r(u)} \left (a_{r(u)} + b_{r(u)} x \right ),
\end{align*}
and using a well-known limit composition theorem: if $\lim\limits_{r\to \infty} f_r = c, \lim\limits_{u \to \infty} r(u) = \infty$ and $f_r$ is continuous then $\lim\limits_{u \to \infty} f_{r(u)} = c$. On the other hand, we find
\begin{align*}
F^{\text{POT}}(x) & = \lim_{u\to \infty} \mathcal{P}_{\text{POT}}(X < u + \tilde{a}_u + \tilde{b}_u x | X> u) \\
& = \lim_{u \to \infty} \mathbf{F}_{r(u)} \left (u + \tilde{a}_{u} + \tilde{b}_{u} x \right ),
\end{align*}
which gives the relations between the constants:
\begin{align}
F^{\text{POT}}(x) = F^{\text{free}}(x) \quad \longrightarrow \quad 
\begin{cases}
u + \tilde{a}_u = a_{r(u)} \\
\tilde{b}_u = b_{r(u)}
 \end{cases}.
\end{align}
As was mentioned before, all three laws in the POT approach are typically expressed in terms of the generalized Pareto distribution $G_\beta(x)$:
\begin{align}
\label{genpareto}
G_{\beta\geq 0}(x) & = 
\begin{cases} 
0, & x<0, \\
1-(1+\beta x)^{-\frac{1}{\beta}}, & x>0, \\
\end{cases} \nonumber \\
G_{\beta< 0}(x) & = 
\begin{cases} 
0, & x<0, \\
1-(1+\beta x)^{-\frac{1}{\beta}}, & x\in \left (0,-\frac{1}{\beta} \right ), \\
1, & x> -\frac{1}{\beta}.
\end{cases}
\end{align}
where $\lim\limits_{\beta\to 0} \left ( 1-(1+\beta x)^{-\frac{1}{\beta}} \right ) = 1 - e^{-x}$.
We find $F^{\text{POT}}_{I}(x) =  G_{0}(x)$, $F^{\text{POT}}_{II}(x) =  G_{1/\gamma}(\gamma(x-1))$ and $F^{\text{POT}}_{III}(x) =  G_{-1/\gamma}(\gamma(x+1))$ as can be seen also in Tab. \ref{tab2}. 

\section{Step functions $T(x),t(x)$}
\label{appconst}
We calculate step functions 
\begin{align*}
T(x) & = \lim\limits_{m\to \infty} \theta\left (f(x_m) - \alpha_m \right ), \\
t({x}) & = \lim_{m\to \infty} \theta(f(x_m)).
\end{align*}
with $x_m = a_m + b_m x$ and constants $a_m,b_m$ given in Tab. \ref{tab1} for cases of Gumbel, Fr\'echet and Weibull. The main tool is an explicit definition of the step function $\theta(a) = 1$ if $a>0$ and $\theta(a) = 0$ when $a<0$. That is, both formulas are rewritten as
\begin{align*}
    T(x) & = \begin{cases} 
    1, & \lim\limits_{m\to \infty} f(a_m + b_m x) > \alpha_m \\
    0, & \lim\limits_{m\to \infty} f(a_m + b_m x) < \alpha_m 
    \end{cases} ,\\
    t(x) & = \begin{cases} 
    1, & \lim\limits_{m\to \infty} f(a_m + b_m x) > 0 \\
    0, & \lim\limits_{m\to \infty} f(a_m + b_m x) < 0 
    \end{cases}.
\end{align*}
\subsection{$T(x)$}
In the following we deal with respective inequalities in all three classical cases. 
\paragraph{Gumbel.} Constants are equal to $a_n = f^{-1}(\alpha_m)$ and $b_n = f^{-1}(\alpha_{me}) - f^{-1}(\alpha_{m})$ so that:
\begin{align*}
     f \left ( f^{-1}(\alpha_m) + \left [ f^{-1}(\alpha_{me}) - f^{-1}(\alpha_{m}) \right ] x \right ) > \alpha_m.
\end{align*}
As function $f$ is increasing and monotonic, so is the inverse $f^{-1}$ and so the condition is equivalent to
\begin{align*}
     f^{-1}(\alpha_m) + \left [ f^{-1}(\alpha_{me}) - f^{-1}(\alpha_{m}) \right ] x > f^{-1} (\alpha_m).
\end{align*}
Additive term $f^{-1}(\alpha_m)$ cancels and moreover we divide out a positive factor $f^{-1}(\alpha_{me}) - f^{-1}(\alpha_{m})>0$ leaving only $x>0$. Same arguments applied to the opposite condition $x<0$ give:
\begin{align*}
    T^{\text{Gumbel}}(x) = \theta(x).
\end{align*}
\paragraph{Fr\'echet.} In this case the inequality reads
\begin{align*}
    f (f^{-1}(\alpha_m)x) > \alpha_m \quad \to \quad x > 1,
\end{align*}
which with the opposite condition gives $T^{\text{Fr\'echet}}(x) = \theta(x-1)$.
\paragraph{Weibull.} Now the step function inequality is given by
\begin{align*}
    f\left (x_+ + (x_+ - f^{-1}(\alpha_m))x \right ) > \alpha_m,
\end{align*}
with $x_+>0$ being the rightmost edge of the spectrum. In particular, $f^{-1}(1) = x_+$ and we have
\begin{align*}
    x_+ (1+x) > (1+x) f^{-1}(\alpha_m),
\end{align*}
which is true then $1+x>0$ since then the term cancels and the remaining $ x_+ > f^{-1}(\alpha_m)$ is always true as $\alpha_m < 1$. Otherwise i.e. when $1+x<0$ the remaining inequality $x_+ < f^{-1}(\alpha_m)$ is false. The same reasoning for the opposite case results in $T^{\text{Weibull}}(x) = \theta(x+1)$.
\subsection{$t(x)$}
Before inspecting inequalities we introduce a regularized $t_\epsilon$ reducing to previous step function as $\lim_{\epsilon\to 0} t_\epsilon(x) = t(x)$. We have
\begin{align*}
        t_\epsilon(x) & = \begin{cases} 
    1, & \lim\limits_{m\to \infty} f(a_m + b_m x) > \epsilon \\
    0, & \lim\limits_{m\to \infty} f(a_m + b_m x) < \epsilon
    \end{cases}.
\end{align*}
\paragraph{Gumbel.} Step function inequality reads
\begin{align*}
     x > \frac{f^{-1} (\epsilon)-f^{-1}(\alpha_m)}{f^{-1}(\alpha_{me}) - f^{-1}(\alpha_{m})}
\end{align*}
which, in the $m \to \infty$ limit $\frac{f^{-1} (\epsilon)-f^{-1}(\alpha_m)}{f^{-1}(\alpha_{me}) - f^{-1}(\alpha_{m})} \to -\infty$ as the denominator is positive and goes to zero while numerator reaches negative infinity. Resulting inequality $ x> -\infty$ is always satisfied and so $t^{\text{Gumbel}}(x) = 1$.
\paragraph{Fr\'echet.} Step function inequality is given by
\begin{align*}
    x > \frac{f^{-1} (\epsilon)}{f^{-1}(\alpha_{m})}
\end{align*}
which, in the $m\to \infty$ limit $\frac{f^{-1} (\epsilon)}{f^{-1}(\alpha_{m})} \to 0$ and so with the opposite inequality we find $t^{\text{Fr\'echet}}(x) = \theta(x)$.
\paragraph{Weibull.} In this case, step function inequality reads
\begin{align*}
    x > \frac{f^{-1} (\epsilon) - x_+ }{x_+ - f^{-1}(\alpha_m)}.
\end{align*}
with $x_+$ being the rightmost end of the spectrum. We know that $x_+ > f^{-1}(\alpha_m)$ while $x_+ - f^{-1}(\alpha_m) \to 0$ in the $m\to \infty$. In the $m\to \infty$ limit the r.h.s. of above inequality reads  $\frac{f^{-1} (\epsilon) - x_+ }{x_+ - f^{-1}(\alpha_m)} \to -\infty$ resulting in $x > -\infty$ which is always satisfied. With the opposite inequality we finally obtain $t^{\text{Weibull}}(x) = 1$.

\end{document}